\RequirePackage{fix-cm}
\documentclass[twocolumn,epjc3]{svjour3}  
\usepackage{lineno}
\RequirePackage{graphicx}
\RequirePackage{physics}
\RequirePackage{latexsym}
\RequirePackage{amssymb}
\RequirePackage[numbers,sort&compress]{natbib}
\RequirePackage{amsfonts}
\RequirePackage{float}
\RequirePackage[colorlinks,citecolor=blue,urlcolor=blue,linkcolor=blue]{hyperref}
\RequirePackage{amsmath}
\RequirePackage[compat=1.1.0]{tikz-feynman}
\usepackage[dvipsnames]{xcolor}
\definecolor{myblue}{cmyk}{0.65, 0.37, 0.0, 0.19}

\usepackage[normalem]{ulem}
\usepackage{orcidlink} 

\journalname{Eur. Phys. J. C}

\begin{document}

\title{\boldmath Fermionic Dark Matter in a Scotogenic Model with a Complex Scalar Singlet}

\author{G. Ardila-Tafurth \orcidlink{0009-0006-8074-4780} \thanksref{e1,addr1}
        \and
        A. Flórez \orcidlink{0000-0002-3222-0249} \thanksref{e2,addr1}
        \and
        V. Martín Lozano \orcidlink{0000-0002-9601-0347}  \thanksref{e3,addr2} \and 
        A. Vicente \orcidlink{0000-0002-1137-4695} \thanksref{e4, addr2, addr3}
         }

\thankstext{e1}{ga.ardila10@uniandes.edu.co}
\thankstext{e2}{ca.florez@uniandes.edu.co}
\thankstext{e3}{victor.lozano@ific.uv.es}
\thankstext{e4}{avelino.vicente@ific.uv.es}

\institute{Departamento de Física, Universidad de los Andes, Bogotá, Colombia  \label{addr1}
           \and
           Instituto de Física Corpuscular (IFIC), CSIC-Universitat de València, Spain\label{addr2},\and Departament de Física Teòrica, Universitat de València,
46100 Burjassot, Spain \label{addr3}
}

\newcommand{\ga}[1]{{\color{red}[GA: #1]}}
\date{\today}
\date{\today}

\maketitle

\begin{abstract}

We study the phenomenology of the complex scoto-singlet model, an extension of the scotogenic framework that simultaneously accounts for radiative neutrino masses and fermionic dark matter. We perform a Markov Chain Monte Carlo scan of the parameter space, imposing theoretical consistency together with constraints from neutrino oscillation data, lepton-flavor violation, collider searches and the observed dark matter relic abundance. Viable dark matter solutions are found both near the Higgs resonance and in coannihilation regions, extending the allowed mass range with respect to the minimal scotogenic model. Direct- and indirect-detection signals are strongly suppressed, placing most of the viable parameter space beyond the reach of current experiments. At colliders, monojet searches offer the best prospects to probe our model in compressed-spectrum scenarios, while long-lived charged scalars may give rise to displaced or detector-stable signatures.

\end{abstract}

\section{Introduction}
\label{introduction}

The observation of neutrino oscillations and the indirect evidence for the existence of dark matter (DM) from astronomical and cosmological phenomena provide clear motivations to explore new physics scenarios beyond the Standard Model (SM). Among the simplest and most predictive frameworks addressing both issues simultaneously is the scotogenic model~\cite{PhysRevD.54.5693, PhysRevD.73.077301}, in which neutrino masses are generated radiatively at one loop by new particles belonging to a dark sector. The same symmetry that forbids tree-level neutrino masses ensures the stability of a DM candidate, establishing a direct connection between neutrino physics and DM. Motivated by this mechanism, a wide variety of extensions of the scotogenic framework have been systematically explored in the literature~\cite{law2013class, Restrepo:2013aga,Escribano:2020iqq}.~\footnote{A comprehensive review on radiative neutrino mass models can be found in~\cite{Cai:2017jrq}, whereas a recent review on neutrino masses sourced by dark matter is given by~\cite{Avila:2025qsc}.} These typically involve additional scalar or fermionic fields, leading to diverse  phenomenological scenarios and, potentially, testable collider signatures.

In this work, we consider a simple extension of the original scotogenic model, referred to as the \textit{complex scoto-singlet model}~\cite{escribano2025exploring}, where an additional complex scalar singlet is added. This extension focuses on charged lepton flavor violation (LFV) processes and on the impact of the extended scalar sector on Higgs observables, in particular the $h \to \mu^+\mu^-$ and $h \to \gamma\gamma$ signal strengths. A distinctive feature of this approach is that the mechanism responsible for neutrino mass generation remains unchanged with respect to the minimal model. As a result, the neutrino sector retains its predictive structure, allowing a richer phenomenology in the DM sector. Despite its minimal field content, the presence of the complex singlet leads to a non-trivial modification of the scalar sector. In particular, it introduces new interactions and mixing effects that can significantly affect DM annihilation channels and co-annihilation processes, which  in turn impact the resulting relic density.

Herein, we perform the first complete phenomenological study of this model, considering all relevant theoretical and experimental constraints. The impact of scalar mixing on the structure of the viable parameter space, especially in regions with compressed spectra or relevant co-annihilation effects, is explored. The analysis includes bounds from DM direct and indirect detection experiments, as well as constraints from collider searches. The feasible parameter space of the model, after considering the restrictions, is obtained through a numerical analysis using a Markov Chain Monte Carlo (MCMC)~\cite{Markov1971} approach that consistently incorporates neutrino oscillation data, together with bounds from LFV processes, collider searches and cosmological measurements of the DM relic abundance. Special attention is placed on ensuring the theoretical consistency of the scalar potential, including boundedness-from-below and perturbativity constraints.

Our study reveals several interesting phenomenological features. The
fermionic DM candidate can reproduce the observed relic abundance near the
Higgs resonance and, away from this region, mainly through scalar
coannihilation for masses above approximately $250$ GeV. The additional interactions associated with the charged scalar also allow viable solutions at lower masses than in the minimal scotogenic model. Furthermore, the predicted spin-independent DM-nucleon scattering cross section remains several orders of magnitude below the current LZ limit~\cite{LZ:2024zvo}, with a significant fraction of the viable parameter space lying below the neutrino floor. Present-day annihilation is dominated by the $\tau^+\tau^-$ and $\mu^\pm\tau^\mp$ final states, although the corresponding signals remain well below the current Fermi-LAT and H.E.S.S. limits from dwarf spheroidal galaxies (dSphs)~\cite{Fermi-LAT:2025gei}. Finally, monojet searches  may  probe part of the low-mass parameter space at the High Luminosity Large Hadron Collider (HL-LHC)

The paper is organized as follows. In Sec.~\ref{sec:mods} we describe the model, its particle content, interactions and mass spectrum. Sec.~\ref{sec:setup} presents the methodological framework used to explore the parameter space, while incorporating all relevant experimental and theoretical constraints to identify the viable regions of parameter space featuring a fermionic DM candidate. The phenomenological results are presented in Sec.~\ref{sec:pheno}, while the conclusions and final remarks are shown in Sec.~\ref{sec:summ}.

\section{The Model}
\label{sec:mods}

The complex scoto-singlet model~\cite{escribano2025exploring} is a simple variant of the scotogenic model which extends the particle content of the SM by means of 3 Majorana-like singlet fermions $\textrm{N}_i ~(i=1,2,3)$, one $\textrm{SU}(2)_L$ scalar doublet $\eta$, and a complex scalar singlet field $\phi$ with non-zero hypercharge.~\footnote{We emphasize that the scalar singlet introduced in our model is electrically charged, and hence complex. For a related model featuring a real scalar singlet we refer to~\cite{Beniwal:2020hjc,Escribano:2023hxj}.} Additionally, a $\mathbb{Z}_2$ symmetry is imposed, under which all the new particles are odd, while the SM particle content remains even as illustrated in Tab~\ref{tab:particlecontent}.
\begin{table}[H]
        \centering
        \begin{tabular}{|c|c|c|c|c|}
        \hline
             Field & Generations & SU(2)$_L$ & U(1)$_Y$ & $\mathbb{Z}_2$  \\\hline
             $\ell_L$& 3& 2& -1/2& 1\\\hline
             $\textrm{e}_R$ &3&1&-1&1\\\hline
             $\textrm{H}$&1&2&1/2&1\\\hline
             $\phi$& 1& 1&-1&-1\\\hline
             $\eta$& 1&2&1/2&-1\\\hline
             $\textrm{N}$ & 3&1&0&-1\\\hline
        \end{tabular}
        \caption{Leptonic and scalar particle content of the scoto-singlet model.}
        \label{tab:particlecontent}
    \end{table}
    Here, $\mathrm{H}$ denotes the usual SM Higgs doublet, while $\mathrm{e}_R$ are the electrically charged right-handed leptons. Moreover, the imposed discrete symmetry stabilizes the lightest of the new $\mathbb{Z}_2$-odd states, which can be taken to be electrically neutral. Therefore, this setup predicts the existence of a potentially viable DM candidate which can be either scalar or fermionic. 

    Given the particle content of the model, the most general Yukawa Lagrangian, compatible with all symmetries, is
    \begin{align}
        -\mathcal{L}\supset y_{ij}\bar{\textrm{N}}_i^c \eta \ell_j+ \kappa_{ij}\bar{\textrm{N}}_i^c \phi^* \textrm{e}_{R,j}+\frac{1}{2}M_{ij} \bar{\textrm{N}}_i^c \textrm{N}_j + \textrm{h.c}, 
        \label{eq:lagrangian}
    \end{align}
where $y_{ij}$, $\kappa_{ij}$, and $M_{ij}$ are $3\times 3$ matrices corresponding to the Yukawa interactions between the SM leptons and the dark-sector scalars, and to the mass matrix of the singlet neutral fermions. Without loss of generality, we take $M_{ij}$ to be diagonal in our calculations.  
Regarding the scalar sector, the most general $\mathbb{Z}_2$ and gauge invariant potential is given by 
\begin{align}
            V&=m_\textrm{H}^2 (\textrm{H}^\dagger \textrm{H})+\frac{\lambda_1}{2}(\textrm{H}^\dagger \textrm{H})^2+m_\eta^2 (\eta^\dagger\eta) +\frac{\lambda_2}{2}(\eta^\dagger\eta)^2 \nonumber \\
                &+\lambda_3(\textrm{H}^\dagger \textrm{H})(\eta^\dagger\eta) +\lambda_4(\textrm{H}^\dagger\eta)(\eta^\dagger \textrm{H})+ \left[\frac{\lambda_5}{2} \, (\textrm{H}^\dagger\eta)^2+\textrm{h.c} \right] \nonumber \\
                &+m_\phi^2 (\phi^*\phi)+\lambda_\phi (\phi^*\phi)^2+\lambda_{\eta\phi}(\eta^\dagger\eta)(\phi^*\phi) \nonumber \\
                &+\lambda_{\textrm{H}\phi}(\textrm{H}^\dagger \textrm{H})(\phi^*\phi)+ \left[\mu \, (\eta^\dagger \tilde{\textrm{H}}\phi^*)+\textrm{h.c} \right] \, ,
            \label{eq:potential}
\end{align}
where $m_\textrm{H}$, $m_\eta$, $m_\phi$ and $\mu$ are parameters with mass dimension, while the rest of the parameters remain dimensionless. 

\subsection{Scalar mass spectrum}
\label{sec:scalarspectrum}
After electroweak symmetry breaking the scalar doublets of the model are parametrized as
\begin{align}
    \textrm{H}=\mqty( \textrm{H}^+\\ \frac{1}{\sqrt{2}}(v+h+iA)), \quad \eta=\mqty(\eta^+\\\frac{1}{\sqrt{2}}(\eta_R+i\eta_I)),
\end{align}
where $v\approx246~\textrm{GeV}$ is the SM vacuum expectation value (VEV). Furthermore,  $\textrm{H}^+$ becomes the longitudinal component of the SM $W^+$ vector boson, $A$ is the Goldstone boson associated with the mass of the SM $Z$ boson, $h$ and $\eta_R$ are the $CP$-even components of the doublets, and $\eta_I$ is a physical pseudoscalar field.

Due to the presence of the $\mu$ term in the scalar potential, the charged component of the $\eta$ doublet and the $\phi$ singlet mix to form two mass eigenstates, $S_1^{\pm}$ and $S_2^\pm$, defined as 
\begin{align}
    \mqty(S_1^{\pm}\\S_2^\pm)=\mqty(\cos\theta & -\sin\theta\\\sin\theta &\cos \theta)\mqty(\eta^\pm\\\phi^\pm).
    \label{eq:mixing_chargedScalars}
\end{align}
 The mixing angle, $\theta$,  is  related to the parameters of the potential via 
 \begin{align}
     \tan(2 \,\theta)=\frac{\sqrt{2} \, v \, \mu}{\widetilde{m}_\eta^2-\widetilde{m}_\phi^2} \, ,
 \end{align}
 with $\widetilde{m}_\eta^2=m_\eta^2+\frac{1}{2}\lambda_3 v^2$ and $\widetilde{m}_\phi^2=m_\phi^2+\frac{1}{2}\lambda_{\textrm{H}\phi}v^2$. Regarding the masses of the new charged states, these are given  by
 \begin{align}
     m_{S_{1,2}}^2=\frac{1}{2}\left[\widetilde{m}_\phi^2+\widetilde{m}_\eta^2\mp \sqrt{(\widetilde{m}_\eta^2
     -\widetilde{m}_\phi^2)^2 +2v^2 \, \mu^2}\right],
     \label{eq:chargedmasses}
 \end{align}
while the masses for the remaining physical states are
 \begin{equation}
            \begin{split}
                &m_{\eta_R}^2=m_\eta^2+\frac{v^2}{2} (\lambda_3+\lambda_4+\lambda_5)\\
                &m_{\eta_I}^2=m_\eta^2+\frac{v^2}{2} (\lambda_3+\lambda_4-\lambda_5)\\
                &m_h^2=\lambda_1 v^2.
                \label{eq:neutralscalarmasses}
            \end{split}
        \end{equation}

\subsection{Neutrino masses}
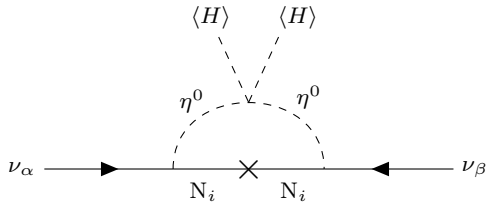
\begin{figure}[H]
\centering
\begin{tikzpicture}
  \begin{feynman}
    
    \vertex (vi) at (-3,0) {$\nu_\alpha$};
    \vertex (v1) at (-1,0);
    \vertex (v2) at ( 1,0);
    \vertex (vj) at ( 3,0) {$\nu_\beta$};

    \diagram*{
      (vi) -- [fermion] (v1),
      (v1) -- [plain] (v2),
      (v2) -- [anti fermion] (vj)
    };

    \node at (-0.6,-0.3) {$\textrm{N}_i$};
    \node at ( 0.6,-0.3) {$\textrm{N}_i$};

    \node at (0,0) {\Large $\times$};

    \diagram*{
      (v1) -- [scalar, half left] (v2)
    };

    \node at (-0.75,0.90) {$\eta^0$};
    \node at ( 0.8,0.95) {$\eta^0$};

    \vertex (mid) at (0,0.88);

    \vertex (phiL) at (-0.4,1.75);
    \vertex (phiR) at ( 0.4,1.75);
    \diagram*{
      (mid) -- [scalar, dashed] (phiL),
      (mid) -- [scalar, dashed] (phiR)
    };

    \node at (-0.50,2.0) {$\langle H \rangle$};
    \node at ( 0.65,2.0) {$\langle H \rangle$};

  \end{feynman}
\end{tikzpicture}
\caption{Feynman diagram that illustrates the one-loop neutrino mass generating mechanism. Here $\langle H \rangle$ denotes the electroweak VEV and $\eta^0$ the real and complex components of the $\eta$ doublet.}
\label{fig:neutrinodiagram}
\end{figure}

Analogously to the minimal scotogenic model, neutrino masses are generated via the one-loop process illustrated in Fig.~\ref{fig:neutrinodiagram}.

After integration, it is possible to find that the neutrino mass matrix becomes
\begin{equation}
    \begin{split}
        \mathcal{M}_{\alpha\beta}=\sum_{i=1}^3 &\frac{y_{i\alpha}y_{\beta i}}{32\pi^2} M_i\bigg[\frac{m_{\eta_R}^2}{M_i^2-m_{\eta_R}^2}\ln\left(\frac{m_{\eta_R}^2}{M_i^2}\right)\\&-\frac{m_{\eta_I}^2}{M_i^2-m_{\eta_I}^2}\ln\left(\frac{m_{\eta_I}^2}{M_i^2}\right)\bigg],
    \end{split}
    \label{eq:neutrinomassmat}
\end{equation}
where $M_i$ is the mass of the $\textrm{N}_i$ fermion, $y_{ij}$ are the Yukawa couplings between the singlet fermions and the SM lepton doublets, while $m_{\eta_R}$ and $m_{\eta_I}$ are the masses of the $CP$-even and $CP$-odd components of the $\eta$ doublet. Additionally, as follows from Eq.~\eqref{eq:neutralscalarmasses}, the observed smallness of neutrino masses can be naturally explained by taking $\lambda_5 \ll 1$. In the limit $\lambda_5 \to 0$, the mass splitting between $\eta_R$ and $\eta_I$ vanishes and lepton number conservation is restored. Consequently, a small value of the quartic coupling $\lambda_5$ is technically natural in the sense of 't Hooft~\cite{tHooft:1979rat}.

\section{Methodological Setup}\label{sec:setup}
The central part of our study consists of analyzing the DM phenomenology of the model, taking $\textrm{N}_1$ as the DM candidate particle. To that end, we first implemented the model in \texttt{SARAH} (version 4.15.1)~\cite{SARAH2010,SARAH2011,SARAH2013,SARAH2014}, which then generates an output that is used by \texttt{SPheno} (version 4.0.5)~\cite{SPheno2003,SPheno2012,spheno4}, to compute the numerical values of the particle masses, low-energy observables, LFV rates and Higgs observables. The DM observables such as the relic density, the spin-independent DM-Nucleon scattering cross-section and the average effective annihilation cross-section times velocity for indirect signals are calculated using \texttt{micrOmegas} (version 6.0.5)~\cite{Alguero:2023zol}. All the theoretical and experimental constraints considered in this section are incorporated in our analysis via an MCMC numerical method~\cite{Markov1971}. To calculate the production cross-sections at the LHC, we used a Universal Feynman Output~\cite{Degrande:2011ua} module for {\tt MadGraph5\_aMC\@NLO} (version 3.1.0)~\cite{Alwall:2011uj}.

\subsection{Constraints}
\label{sec:constraints}

In this work we considered several theoretical and experimental constraints which impact the parameter space of the model. 

\subsubsection{Unitarity, perturbativity and boundedness from below}

We require the scalar potential to be bounded from below to ensure the stability of the electroweak vacuum, perturbativity of all couplings to guarantee the validity of the perturbative expansion, and tree-level unitarity of scattering amplitudes~\cite{Kannike:2012pe, Kadastik:2009cu}. 

First, unitarity and perturbativity are ensured by imposing that the quartic couplings and the Yukawas must verify
\begin{align*}
        \abs{\lambda_i}<4\pi;\quad \abs{y_{ij}}<\sqrt{4\pi}.
    \end{align*}

Secondly, the boundedness of the potential can be achieved if we require coposititivity conditions on the quartic terms of the potential. These conditions can be written as~\cite{escribano2025exploring}
\begin{align}
    0\leq \lambda_1,\lambda_2, \lambda_\phi;\quad0\leq c_1, c_2,c_3, c_4,
\end{align}
where $c_1$, $c_2$, and $c_3$ correspond to the two-field conditions,
\begin{equation}
\begin{aligned}
c_1 &=
\begin{cases}
\lambda_3+\sqrt{\lambda_1\lambda_2}, 
& \lambda_4-|\lambda_5|\geq 0,\\
\lambda_3+\lambda_4-|\lambda_5|+\sqrt{\lambda_1\lambda_2}, 
& \lambda_4-|\lambda_5|<0,
\end{cases} \\
c_2 &= \lambda_{\textrm{H}\phi}+\sqrt{\lambda_1\lambda_\phi}, \\
c_3 &= \lambda_{\eta\phi}+\sqrt{\lambda_2\lambda_\phi}.
\end{aligned}
\end{equation}
Defining
\begin{equation}
\tilde{\lambda}_3=
\begin{cases}
\lambda_3, & \lambda_4-|\lambda_5|\geq 0,\\
\lambda_3+\lambda_4-|\lambda_5|, & \lambda_4-|\lambda_5|<0,
\end{cases}
\end{equation}
allows us to write the three-field copositivity condition as
\begin{equation}
\begin{aligned}
c_4={}&\sqrt{\lambda_1\lambda_2\lambda_\phi}
+\tilde{\lambda}_3\sqrt{\lambda_\phi}
+\lambda_{\mathrm{H}\phi}\sqrt{\lambda_2} \\
&+\lambda_{\eta\phi}\sqrt{\lambda_1}
+\sqrt{2c_1c_2c_3}.
\end{aligned}
\end{equation}

\subsubsection{Higgs boson mass}

We require the Higgs boson mass to lie within a $\pm 3~\textrm{GeV}$ interval around the measured value~\cite{ParticleDataGroup:2024cfk},
\begin{align}
    m_h \in [122,\,128]~\textrm{GeV},
\end{align}
thereby accounting for theoretical uncertainties in the mass prediction, including higher-order corrections and numerical effects in the spectrum calculation.

\subsubsection{Radiative stability of the Higgs mass}

Due to the presence of the $\mu$ term in the scalar potential, the Higgs mass receives a one-loop correction proportional to $\mu^2$~\cite{Beniwal:2020hjc,escribano2025exploring},
\begin{align}
    \Delta m_{H}^2 \propto \mu^2.
\end{align}
Requiring that the physical Higgs mass remains compatible with the experimentally measured value implies that the $\mu$ parameter should not introduce large radiative corrections. This translates into the condition
\begin{align}
    |\mu| \lesssim 4\pi\,\epsilon\, m_h,
\end{align}
where $m_h \simeq 125~\textrm{GeV}$ is the Higgs boson mass, and $\epsilon \in [0,1]$ parametrizes the level of tolerated fine-tuning. In this work, we take $\epsilon = 1$, which implies $|\mu| \lesssim 1.5~\textrm{TeV}$, unless stated otherwise.

\subsubsection{Electroweak precision observables}
Electroweak precision observables are taken into account through the Peskin--Takeuchi parameters $S$, $T$, and $U$, which encode new physics contributions to gauge boson vacuum polarizations~\cite{Peskin:1991sw}. In models with additional scalar multiplets, these parameters receive one-loop corrections determined by the mass spectrum and electroweak quantum numbers of the new states. We compute these contributions and require consistency with the bounds~\cite{ParticleDataGroup:2024cfk}
\begin{align}
\begin{aligned}   &S=-0.04\pm0.10 \, ,\\&T=0.01\pm 0.12 \, , \\&U=-0.01\pm0.09 \, .
\end{aligned}
\end{align}

Among these observables, the $T$ parameter plays a particularly important
role in the present model, since it is sensitive to custodial-symmetry
breaking induced by the mass splittings between the charged scalars and
the neutral components of the inert doublet. Using the mixing defined in Eq.~\eqref{eq:mixing_chargedScalars}, the dominant beyond the SM (BSM) contributions to the $T$ parameter can be written as~\cite{Grimus:2008nb}

\begin{align}
\Delta T \simeq
\frac{1}{8\pi s_W^2m_W^2}
\Bigg\{&
c^2_\theta \, F\left(m_{S_1}^2,m_0^2\right)
+
s^2_\theta \, F\left(m_{S_2}^2,m_0^2\right)
\nonumber\\
&-
\frac{1}{4}s^2_{2\theta}
F\left(m_{S_1}^2,m_{S_2}^2\right)
\Bigg\},
\label{eq:Tparameter}
\end{align}

where $c_\theta=\cos\theta$, $s_\theta=\sin\theta$, $s_{2\theta} = \sin 2 \theta$, $s_W = \sin \theta_W$ is the sine of the weak mixing angle $\theta_W$, and 
\begin{equation}
F(x,y)
=
\frac{x+y}{2}
-
\frac{xy}{x-y}
\log\left(\frac{x}{y}\right),
\quad
F(x,x)=0 \, .
\end{equation}
In Eq.~\eqref{eq:Tparameter}, the contribution induced by the mass splitting between $\eta_R$ and $\eta_I$ has been neglected, as these states are nearly degenerate. We therefore define $m_0^2 \approx m_{\eta_R}^2 \approx m_{\eta_I}^2$. Finally, the Higgs contributions have been omitted, as they are identical to those of the SM.

As will be shown in Sec.~\ref{sec:DY_VBF}, the charged-scalar mixing angle $\theta$ is small throughout the parameter space considered in this work. Therefore, $c_\theta^2 \approx 1$, $s_\theta^2 \ll 1$, $S_1^\pm \approx \eta^\pm$ and $S_2^\pm \approx \phi^\pm$. In this limit, $\Delta T$ is primarily determined by the mass splitting between $S_1^\pm$ and $\eta_{R,I}$, although sizable contributions from $S_2^\pm$ may also arise if $m_{S_2}^2 \gg m_{S_1}^2$.

\subsubsection{Neutrino oscillation data}
All the points must comply with the experimental constraints from neutrino oscillation data. To include this information, we implement a modified Casas-Ibarra of the Yukawa couplings~\cite{Casas:2001sr,Toma:2013zsa,Cordero-Carrion:2018xre,Cordero-Carrion:2019qtu}
\begin{align}
    y=\sqrt{\Lambda^{-1}} \, R \, \sqrt{\hat{M_\nu}} \, U_{\rm PMNS}^\dagger,
\end{align}
where $R$ is an orthogonal $3\times3$ matrix which we take to be real for simplicity, $\hat{M}_\nu=\textrm{Diag}(m_1,m_2,m_3)$ is a diagonal matrix containing the light neutrino masses, $U_{\rm PMNS}$ is the Pontecorvo-Maki-Nakagawa-Sakata (PMNS) leptonic mixing matrix, and $\Lambda$ is defined as

\begin{equation}
    \begin{aligned}
        &\Lambda=\textrm{Diag}(\Lambda_i);\\
        &\Lambda_i=\frac{M_i}{32\pi^2}\bigg[\frac{m_{\eta_R}^2}{M_i^2-m_{\eta_R}^2}\ln\left(\frac{m_{\eta_R}^2}{M_i^2}\right)\\&-\frac{m_{\eta_I}^2}{M_i^2-m_{\eta_I}^2}\ln\left(\frac{m_{\eta_I}^2}{M_i^2}\right)\bigg].
    \end{aligned}
\end{equation}
For this analysis, we considered the values of the angles of the PMNS matrix, as well as the mass-squared differences, within the experimental $3\sigma$ range~\cite{Esteban:2020cvm, deSalas:2020pgw}, while the Majorana phases are allowed to freely vary from 0 to $2\pi$.

\subsubsection{Absolute neutrino mass scale}
Cosmological observations provide stringent constraints on the sum of the active neutrino masses. Using measurements of the cosmic microwave background from the Planck collaboration~\cite{aghanim2020planck} in combination with baryon acoustic oscillation (BAO) data from DESI~\cite{adame2025desi}, one obtains an upper bound of $\sum m_\nu < 0.115~\mathrm{eV}$ at 95\% C.L. In this work, we adopt this result as an upper limit.

\subsubsection{Lepton flavor violation}
LFV processes provide strong constraints on the Yukawa structure of scotogenic-like models~\cite{Toma:2013zsa}. In the present framework, LFV arises at the one-loop level through the exchange of the new scalar and fermionic states, leading to processes such as $\mu \to e \gamma$. The corresponding branching ratios depend on the Yukawa couplings and the masses of the charged scalars running in the loop~\cite{escribano2025exploring}. We impose the current experimental limits on LFV observables, in particular the bound on $\mathrm{BR}(\mu \to e \gamma)$, to further restrict the viable parameter space using the most recent result from the MEG-II collaboration~\cite{afanaciev2025new}
\begin{align}
    \mathrm{BR}(\mu\to e\gamma)<1.5\times 10^{-13}\quad @~90\mathrm{\% ~C.L}.
\end{align}
\subsubsection{Higgs observables}

We also incorporate constraints derived from Higgs observables. In the present model, the scalar potential in Eq.~\eqref{eq:potential} contains interactions that couple the Higgs boson to additional neutral states. After electroweak symmetry breaking, these terms induce trilinear couplings between the Higgs and the new scalar fields, allowing for decays of the form $h \to \eta^0 \eta^0$ at tree level or $h \to \textrm{N}_1 \textrm{N}_1$ at the one-loop level, whenever kinematically allowed. These processes contribute to the invisible decay width of the Higgs boson and are tightly constrained by LHC measurements of its invisible branching ratio. Hence, we impose the bound from Ref.~\cite{ParticleDataGroup:2024cfk},
\begin{align}
    \mathrm{BR}(h \to \mathrm{inv}) < 0.105,
\end{align}
at 95\% C.L.

In addition, the presence of new scalar states modifies the Higgs signal strengths in visible channels. In particular, the $h \to \gamma\gamma$ decay receives additional loop contributions from the charged scalars, while the $h \to \mu^+ \mu^-$ channel probes modifications to the Higgs couplings to leptons. Assuming that the Higgs production mechanism remains SM-like, we define the signal strengths as
\begin{align}
    R_{XX} = \frac{\sigma(pp \to h)}{\sigma(pp \to h)_{\mathrm{SM}}}
    \frac{\mathrm{BR}(h \to XX)}{\mathrm{BR}(h \to XX)_{\mathrm{SM}}}.
\end{align}

We impose the combined experimental constraints on the signal strengths from both ATLAS~\cite{ATLAS:2020fzp,ATLAS:2022tnm} and CMS~\cite{CMS:2020xwi,CMS:2021kom,CMS:2022dwd},
\begin{align}
    R_{\mu\mu} \in [0.69,\,1.7], \qquad
    R_{\gamma\gamma} \in [0.99,\,1.17],
\end{align}
at 68\% and 95\% C.L., respectively.

\subsubsection{Collider constraints from LEP}

Collider searches at LEP impose important constraints on the masses of new electroweak states. In particular, the absence of signals for pair production of charged and neutral scalars via electroweak interactions sets lower bounds on their masses and restricts kinematically accessible regions. These constraints arise from processes mediated by the $Z$ and $W$ bosons.

In this work, we impose the following conditions ~\cite{ALEPH:2001oot, DELPHI:2003uqw, L3:2003fyi, OPAL:2003nhx}:
\begin{equation}
\begin{aligned}
    &m_{S_i} > 80~\mathrm{GeV}, \\
    &2 m_{S_i} > m_Z, \quad m_{S_1} + m_{S_2} > m_Z, \quad m_{\eta_R} + m_{\eta_I} > m_Z, \\
    &m_{S_i} + m_{\eta_R}> m_W, \quad m_{S_i} + m_{\eta_I} > m_W,
\end{aligned}
\end{equation}
where $S_i$ denotes the charged scalar states. 
These conditions ensure consistency with LEP limits on scalar pair production through electroweak interactions, preventing on-shell production via $Z$- and $W$-mediated processes.

\subsection{Numerical method}
Our analysis follows Refs.~\cite{Sarazin:2021nwo, deNoyers:2024qjz, Ardila-Tafurth:2025qdf, Alvarez:2023dzz} by implementing a Metropolis--Hastings algorithm~\cite{Metropolis:1953am, Hastings:1970aa} in \texttt{Python}, enabling an efficient exploration of the parameter space by preferentially sampling regions compatible with the imposed experimental constraints, encoded in an effective likelihood function.

First, we define an effective likelihood associated with each observable as
\begin{align}
    \ln\mathcal{L}_i = -\frac{\left|\theta^{i}_{\mathrm{obs}} - \theta^{i}_{\mathrm{exp}}\right|^2}{2 \, \sigma_i^2},
\end{align}
where $\theta^{i}_{\mathrm{obs}}$ denotes the value of the observable computed at a given point in the scan, $\theta^{i}_{\mathrm{exp}}$ is the corresponding experimental value, and $\sigma_i$ its associated uncertainty, as described in Sec.~\ref{sec:constraints}. The total likelihood is then given by $\mathcal{L} = \prod_i \mathcal{L}_i$, assuming that all constraints are uncorrelated.

For observables subject to upper limits, we adopt a one-sided Gaussian likelihood, taking a width corresponding to $10\%$ of the experimental bound. In these cases, if the predicted value lies below the experimental limit, the likelihood is set to unity, while values above the bound are penalized according to the Gaussian profile.
The values and ranges of the parameters used in the scan are summarized in Table~\ref{tab:scan}.
\begin{table}[H]
    \centering
    \begin{tabular}{|c|c|}
    \hline
         Parameter & Range  \\
         \hline
         $m_\eta^2$&$[2.5\times10^3,3\times10^7]$\\\hline
         $m_\phi^2$&$[2.5\times10^3,3\times10^7]$\\\hline
         $\lambda_2$, $\lambda_3$, $\lambda_4$ & $[0,4\pi]$\\
         \hline
         $\lambda_5$&$[1\times10^{-9},1]$\\\hline
         $\lambda_\phi$, $\lambda_{\eta\phi}$, $\lambda_{\mathrm{H}\phi}$&$[0,4\pi]$\\\hline
         $\mu$&$[0, 1500]$\\\hline
         $M_{\mathrm{\textrm{N}_1}}$&$[50,5000]$\\\hline
         $M_{\mathrm{N}_2}$, $M_{\mathrm{N}_3}$&$[100,5000]$\\\hline
         $m_{\nu_1}$&$[10^{-32}, 10^{-11}]$\\
         \hline
    \end{tabular}
    \label{tab:scan}
    \caption{Values of the selected input parameters considered in the scan. The mass parameters are expressed in \textrm{GeV}, while all other parameters are dimensionless.}
\end{table}

In our numerical scan, we have restricted all quartic couplings to positive values, even though the copositivity conditions would allow some of them to take negative values. In particular, we have considered only $\lambda_4>0$. Extending the analysis to $\lambda_4<0$ does not qualitatively modify our conclusions. Although in this case the charged scalar $S_1^\pm$ becomes heavier than the neutral scalars, $S_2^\pm$ becomes even heavier, thereby increasing the scalar mass splittings. This, in turn, substantially increases the BSM contribution to the $T$ parameter, $\Delta T$, leaving only a very limited viable region of parameter space in which the two-body decay channel $S_1^\pm\to\eta_{R/I}\,W^{\pm}$ is never kinematically accessible. Finally,  for the elements of the $\kappa$ matrix, we adopt the parametrization
\begin{align}
    \kappa_{ij} = \abs{\kappa_{ij}} e^{i\theta_{ij}},
\end{align}
where the moduli $\abs{\kappa_{ij}}$ are scanned over the interval $[0,\,4\pi]$, and the phases $\theta_{ij}$ over $[0,\,2\pi]$, as in~\cite{escribano2025exploring}.

After selecting the points that satisfy all imposed theoretical and experimental constraints, we compute the relevant dark matter observables, namely the relic density, the spin-independent DM--nucleon scattering cross section, and the thermally averaged annihilation cross section for indirect detection. The resulting viable configurations are then used to derive the phenomenological predictions presented in Sec.~\ref{sec:pheno}.

In our analysis, we consider two representative mass-spectrum scenarios. The first corresponds to a generic, non-compressed spectrum, where no specific requirement is imposed on the mass splitting between the dark matter candidate and the other $\mathbb{Z}_2$-odd states. The second corresponds to a \textit{compressed mass spectrum} between $\textrm{N}_1$ and $\eta_I$, defined by $\Delta m=\abs{m_{\textrm{N}_1}-m_{\eta_I}}$, for which we consider the benchmark values $\Delta m \leq 30~\mathrm{GeV}$ and $\Delta m \leq 50~\mathrm{GeV}$.

The compressed regime enhances the role of co-annihilations in the early universe through direct or indirect processes, as shown in Refs.~\cite{Ardila-Tafurth:2025qdf, Profumo:2006bx}. Consequently, the relic abundance can be either increased or reduced, depending on the nature of the DM candidate, compared to scenarios in which no requirement on $\Delta m$ is imposed. Such compressed mass-spectrum scenarios have been widely studied in the context of LHC searches for supersymmetric (SUSY) and other DM models~\cite{Aad2020, Sirunyan2019,VBFPhysRevD.87.035029, VBFPhysRevD.90.095022, VBFPhysRevD.91.055025, VBFPhysRevLett.111.061801, natalia2021, CMS:2019zmn, CMS:2016ucr, CMS:2015jsu, Cardona2022, qureshi2024probing, PhysRevD.107.115026, Agin:2024yfs, CMS:2024gyw}.
\section{Phenomenological Analysis}
\label{sec:pheno}

In this section we discuss our results, based on the methodology outlined in Sec.~\ref{sec:setup}, considering $\textrm{N}_1$ as the DM candidate particle.
\subsection{Dark matter phenomenology}
\label{sec:dmpheno}
\subsubsection{Relic abundance}
\begin{figure}
    \centering
    \includegraphics[width=1\columnwidth]{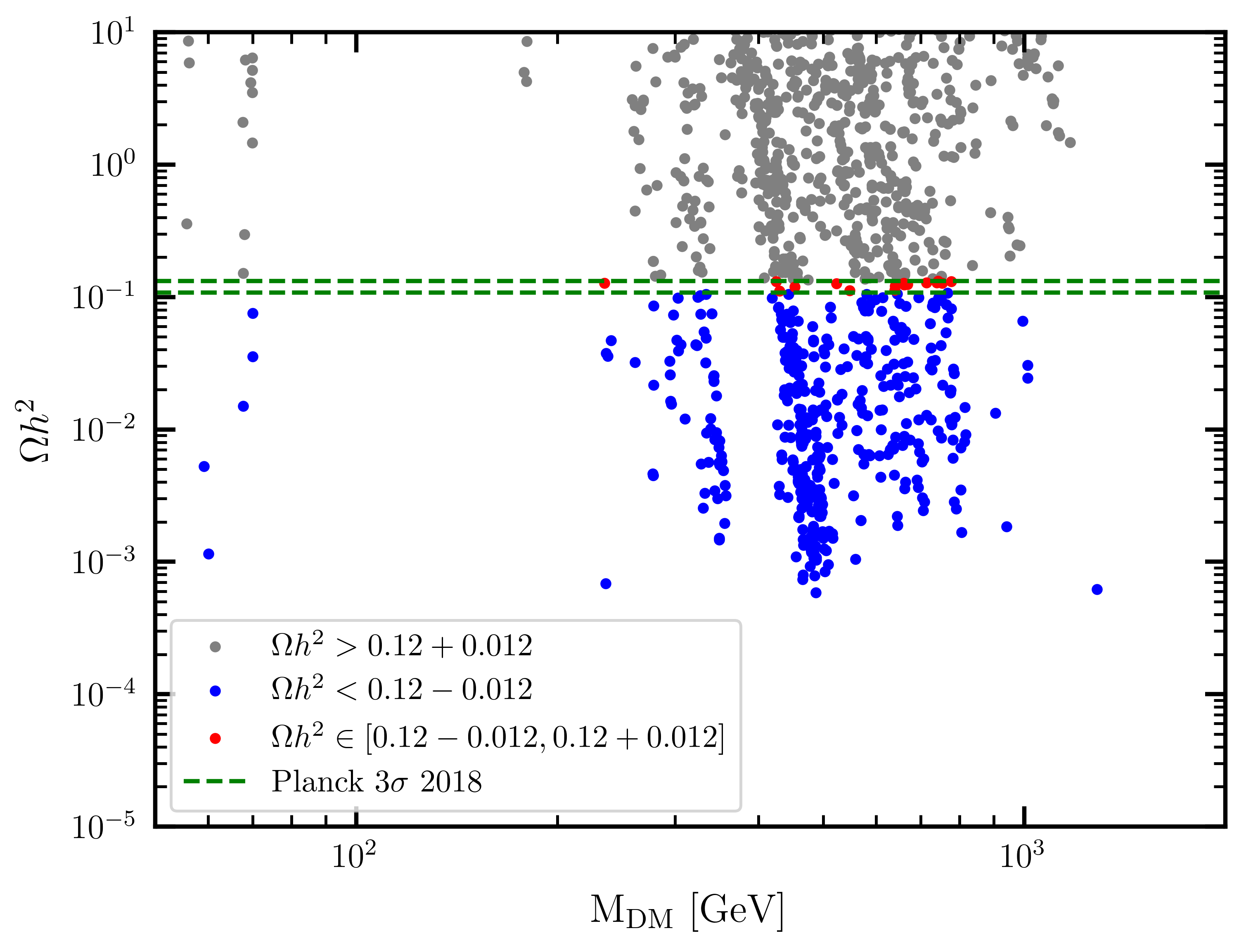}
    \caption{Relic density of $\textrm{N}_1$ as a function of its mass. Red points denote solutions compatible within the $3\sigma$ Planck band. Gray points correspond to overabundant solutions, while blue points indicate underabundant DM scenarios.}
    \label{fig:relic_nodelta}
\end{figure}

First, we study the behavior of the relic DM abundance of the model, which is computed using \texttt{micrOMEGAs} within the MCMC framework. The results are shown in Fig.~\ref{fig:relic_nodelta}, where we present the relic density as a function of the DM mass. The colored regions indicate the compatibility with the observed DM abundance, $\Omega_{\textrm{DM}}h^2=0.12\pm0.036$, where we adopt an enlarged uncertainty to account for numerical and theoretical uncertainties in our calculations. Points lying outside the displayed vertical range are not shown and correspond exclusively to configurations with larger relic-density values, i.e. to increasingly overabundant DM scenarios.

From Fig.~\ref{fig:relic_nodelta} we observe that most of the parameter space leads to an overabundant DM relic density. Nevertheless, viable solutions consistent with the observed value can be identified in specific regions, either within the experimental band or in underabundant scenarios which are still physically viable in models where $\textrm{N}_1$ only accounts for a fraction of the observed abundance and are used in our results. In particular, we find:

\begin{itemize}
    \item $M_{\textrm{N}_1}\approx \frac{m_{h}}{2}$: In this regime, the annihilation of $\textrm{N}_1$ is resonantly enhanced via an $s$-channel Higgs boson, leading to an efficient depletion of the DM abundance. This region is commonly referred to as the \textit{Higgs funnel}. 
    \item $M_{\textrm{N}_1}\gtrsim 250~\textrm{GeV}$: Away from the Higgs funnel, viable solutions are mainly enabled by coannihilation effects involving scalar states close in mass to $\textrm{N}_1$. In particular, the efficient gauge-mediated annihilation channels $S_i^\pm S_j^\mp\rightarrow VV$ and $\eta^0\eta^0\rightarrow VV$, where $V=W,Z$, enhance the effective annihilation cross section and sufficiently reduce the DM abundance.
    
    \item Lower DM masses: Interestingly, the model admits viable solutions at DM masses significantly below those of the minimal scotogenic scenario. This behavior is driven by the additional interactions in the model, namely the $\kappa$ term in Eq.~\eqref{eq:lagrangian}, which increase the annihilation efficiency and relax the lower bound on the DM mass to values of $\mathcal{O}(100~\textrm{GeV)}$.
\end{itemize}

This is possible because $\kappa$, unlike $y$, is not fixed by neutrino oscillation data, as it does not enter the one-loop neutrino mass matrix of Eq.~\eqref{eq:neutrinomassmat}. The Yukawa coupling $y$, on the other hand, is driven to small values by the interplay between the measured PMNS mixing, which prevents $y$ from decoupling from the electron sector, and the stringent bound on $\mathrm{BR}(\mu\to e\gamma)$. The coupling $\kappa$ can instead remain of $\mathcal{O}(1)$, provided its flavor structure is aligned along the $\mu$--$\tau$ directions, thereby evading the most stringent LFV constraints while still efficiently depleting the relic abundance of $\mathrm{N}_1$. The dominance of the $\tau^+\tau^-$ and $\mu^\pm\tau^\mp$ annihilation channels found in our scan is a direct reflection of this structure. It is precisely this freedom that allows viable solutions with the observed relic density to be obtained despite the smallness of $y$, a feature absent in the minimal scotogenic model, where the same Yukawa coupling controls both neutrino masses and dark matter annihilation.

To better understand the origin of the viable relic density solutions, we analyze the dominant processes contributing to the DM annihilation cross section. 

For overabundant points, the relic density is primarily determined by $\textrm{N}_1$ self-annihilations into SM particles, together with subleading contributions from coannihilation processes involving the scalars $S_i^\pm$ and $\eta^0$. Representative channels include $S_i^\pm\,\textrm{N}_1 \to \ell^\pm h$ and $\eta^0\,\textrm{N}_1 \to \nu h$, as well as $S_i^\pm\,\textrm{N}_1 \to \ell^\pm V$, where $V$ denotes a SM vector boson. However, these processes are not efficient enough to sufficiently deplete the DM abundance.

In contrast, the viable solutions compatible with the observed relic density can be divided into two distinct regimes. In the Higgs funnel region, the dominant contribution arises from processes mediated by an $s$-channel Higgs boson, leading to a resonant enhancement of the annihilation cross section into SM final states. 

Away from the funnel, the dominant contribution is instead driven by scalar annihilation channels, particularly $S^\pm_i S^\mp_j \to VV$ and $\eta^0\eta^0\to VV$, which significantly enhance the effective annihilation cross section through coannihilation effects, allowing for a viable relic density. This well-known mechanism~\cite{Profumo:2006bx} has been found to operate in other scotogenic scenarios~\cite{Vicente:2014wga, DeRomeri:2021yjo, DeRomeri:2022cem, escribano2025exploring, Ardila-Tafurth:2025qdf}. 

In the compressed mass spectrum regime, where the mass splitting between $\textrm{N}_1$ and the $\eta_I$ states is small, co-annihilation processes become essential in determining the relic density. For this part of the analysis, we considered an upper threshold of $\Delta m \leq 50~\textrm{GeV}$, following the collider-motivated discussion of Ref.~\cite{Ardila-Tafurth:2025qdf}. In this case, the proximity in mass enhances the thermal population of the scalar states during freeze-out, significantly increasing the efficiency of DM depletion. As a consequence, the relic abundance of $\textrm{N}_1$ decreases with respect to the non-compressed scenario. The dominant contributions arise from co-annihilation channels involving $\eta_I$ and $S_i^\pm$, particularly processes of the form $\eta_I + S_i^\pm \rightarrow VV$, as well as from the self-annihilation of the charged scalars through channels such as $S_i^\pm S_i^\mp \rightarrow VV$ and $S_i^\pm S_i^\mp \rightarrow hh$. These processes are considerably more efficient than the pure $\textrm{N}_1$ annihilation channels, leading to a substantial reduction of the final DM relic density and allowing viable regions of parameter space for lighter fermionic DM masses. These results are not significantly affected by changing the $\Delta m$ value, as the contribution channels to the relic density remain the same for smaller  mass splittings. Hence, for the remaining part of the discussion on DM phenomenology, we restrict our focus to only $\Delta m\leq 50~\rm GeV.$

\subsubsection{Direct detection signatures}

We now discuss the results for $\textrm{N}_1$ direct detection. Since the $\kappa$ and $y$ terms in the Lagrangian do not induce a tree-level contribution to the spin-independent DM--nucleon scattering cross section, we follow Ref.~\cite{Ibarra:2016dlb} and consider the leading loop-induced contribution, given by
\begin{align}
\sigma_{SI}=\frac{4\mu^2 m_p^2 f_p^2}{\pi}\,\Sigma^2,
\end{align}
where $m_p$ is the proton mass, $f_p\approx 0.3$ is the proton scalar form factor~\cite{CMS:2016dhk,ATLAS:2015ciy}, and $\mu=\frac{m_{\textrm{N}_1} m_p}{m_{\textrm{N}_1}+m_p}$ is the reduced mass. The loop contribution $\Sigma$ is given by
\begin{equation} \begin{aligned} \Sigma=&-\frac{y_1^2}{16\pi^2 m_{h}^2 M_{\textrm{N}_1}}\Bigg[\frac{1}{2}(\lambda_3+\lambda_4) \,\mathcal{G}_1\left(\frac{M_{\textrm{N}_1}^2}{m_{\eta^0}^2}\right)\\+&\lambda_3\sin^2\theta \, \mathcal{G }_1\left(\frac{M_{\textrm{N}_1}^2}{m_{S_1}^2}\right) +\lambda_3\cos^2\theta \, \mathcal{G}_1\left(\frac{M_{\textrm{N}_1}^2}{m_{S_2}^2}\right)\\+&\lambda_3\sin^2(2\theta) \,\mathcal{G}_{12}\left(\frac{M_{\textrm{N}_1}^2}{m_{S_1}^2}, \frac{M_{\textrm{N}_1}^2}{m_{S_2}^2}\right)\Bigg], \end{aligned}
\label{eq:SI_formula}
\end{equation}
where $\theta$ is the charged-scalar mixing angle, $y_1=\sqrt{\sum_{j=1}^3 \abs{y_{1j}}^2}$, and the loop functions are defined as
\begin{align}
\mathcal{G}_1(x) &= \frac{x+(1-x)\ln(1-x)}{x}, \\
\mathcal{G}_{12}(x_1,x_2) &= \frac{x_1 \mathcal{G}_1(x_1)-x_2 \mathcal{G}_1(x_2)}{x_1-x_2}.
\end{align}

\begin{figure}[H]
    \centering
    \includegraphics[width=1\columnwidth]{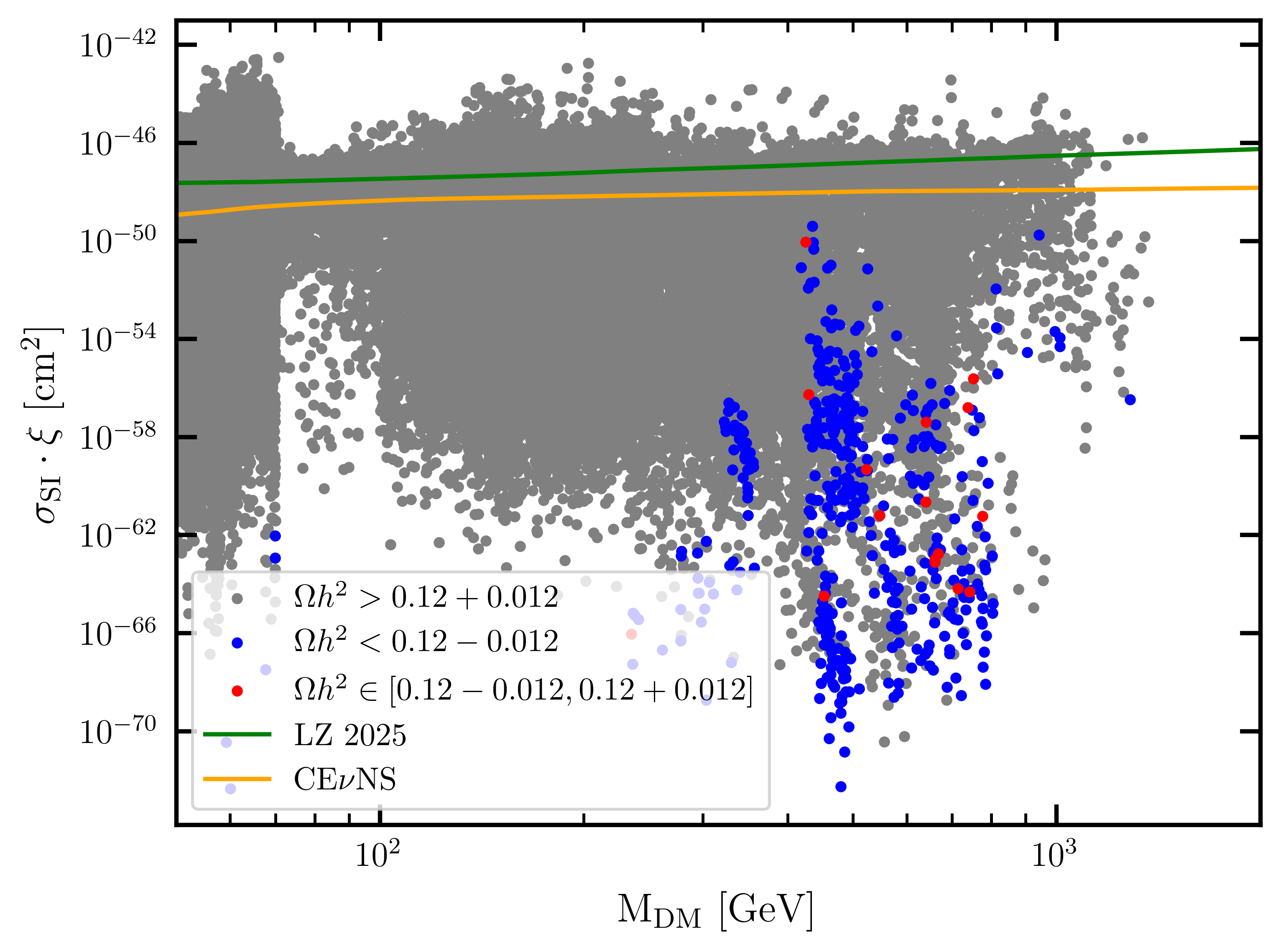}
    \caption{Spin-independent DM--nucleon scattering cross section, $\sigma_{\rm SI}\cdot \xi$, as a function of the DM mass. Gray points correspond to overabundant solutions, blue points to underabundant scenarios, and red points denote parameter space consistent with the observed relic density within $3\sigma$. The solid green line shows the current exclusion limits from LZ, while the orange line indicates the neutrino floor, representing the irreducible background from coherent neutrino--nucleus scattering. All viable solutions lie well below current experimental sensitivities, with a significant fraction of the parameter space falling below the neutrino floor.}
    \label{fig:dd_results}
\end{figure}

After computing Eq.~\eqref{eq:SI_formula}, we show the corresponding results for the spin-independent DM--nucleon scattering cross section in Fig.~\ref{fig:dd_results}. The cross sections shown in this figure are weighted by the relative DM relic density, defined as
\begin{align}
    \xi=\frac{\Omega_{i}}{\Omega_{exp}},
\end{align}
where $\Omega_i h^2$ is the relic density of each point in the scan and $\Omega_{exp}h^2=0.12$ is the experimental central value measured by Planck \cite{aghanim2020planck}. Fig.~\ref{fig:dd_results} also includes the current exclusion limits from the LUX-ZEPLIN experiment~\cite{LZ:2024zvo} (green line), together with the zero-background limit associated with the neutrino floor, which arises from the irreducible background induced by coherent neutrino--nucleus scattering and sets a fundamental limit on the sensitivity of future direct detection searches~\cite{Billard:2013qya}.

The results shown in Fig.~\ref{fig:dd_results} indicate that the predicted spin-independent DM--nucleon scattering cross section lies several orders of magnitude below current experimental limits from LZ. This suppression is a direct consequence of the loop-induced nature of the interaction, together with the typically small Yukawa couplings required to satisfy relic density and LFV constraints.

Furthermore, a significant fraction of the viable parameter space lies below the neutrino floor, particularly in the intermediate and high-mass regions. This implies that probing these scenarios through conventional direct detection experiments will be extremely challenging, as the signal is masked by the irreducible background from coherent neutrino--nucleus scattering. Hence, collider searches and indirect detection experiments become essential complementary probes for this scotogenic scenario.

When imposing the compressed spectrum condition, $\Delta m \leq 50~\textrm{GeV}$, the spin-independent DM--nucleon scattering cross section remains extremely suppressed, with the viable relic density points still lying several orders of magnitude below both the current LZ bounds, falling into the neutrino floor. Compared to the scenario without an upper threshold on $\Delta m$, no significant enhancement of $\sigma_{\rm SI}$ is observed, indicating that the compressed spectrum does not substantially modify the direct detection prospects of the model. Therefore, direct detection constraints continue to have a very limited impact on the fermionic DM parameter space
\subsubsection{Indirect detection}
Given the suppression of the direct detection cross section across the viable parameter space, indirect detection becomes a particularly relevant complementary probe. We therefore analyze the present-day DM annihilation cross section and compare it with current bounds from Fermi-LAT and H.E.S.S. from dSphs~\cite{Fermi-LAT:2025gei}. 

The indirect detection constraints from these experiments are typically derived assuming annihilation into single SM final states. In order to apply these bounds to our scenario, where multiple annihilation channels contribute, we combine the corresponding limits weighted by the branching ratios using a harmonic approximation, given by
\begin{align}
    \frac{1}{\langle \sigma v \rangle_{\textrm{limit}}}
    =
    \sum_i \frac{\textrm{BR}_i}{\langle \sigma v \rangle_i^{\textrm{limit}}}.
    \label{eq:id_approx}
\end{align}
This approach provides an approximate and conservative estimate of the resulting constraints. A more accurate treatment would require a full likelihood analysis including the complete annihilation spectrum, which is beyond the scope of this work. For a detailed discussion, see Ref.~\cite{Armand:2022sjf}.

In our analysis, DM annihilation is dominated by leptonic final states. In particular, the $\tau^+\tau^-$ and $\mu^\pm \tau^\mp$ channels provide the leading contributions, with average branching fractions of $23.2\%$ and $40.1\%$, respectively. Since experimental limits are not provided for the $\mu^\pm \tau^\mp$ final state, we approximate its contribution using the bounds derived for the $\tau^+\tau^-$ and $\mu^+\mu^-$ channels. These limits are then combined according to the corresponding branching ratios within Eq.~\eqref{eq:id_approx}.

\begin{figure}[H]
    \centering
    \includegraphics[width=1\columnwidth]{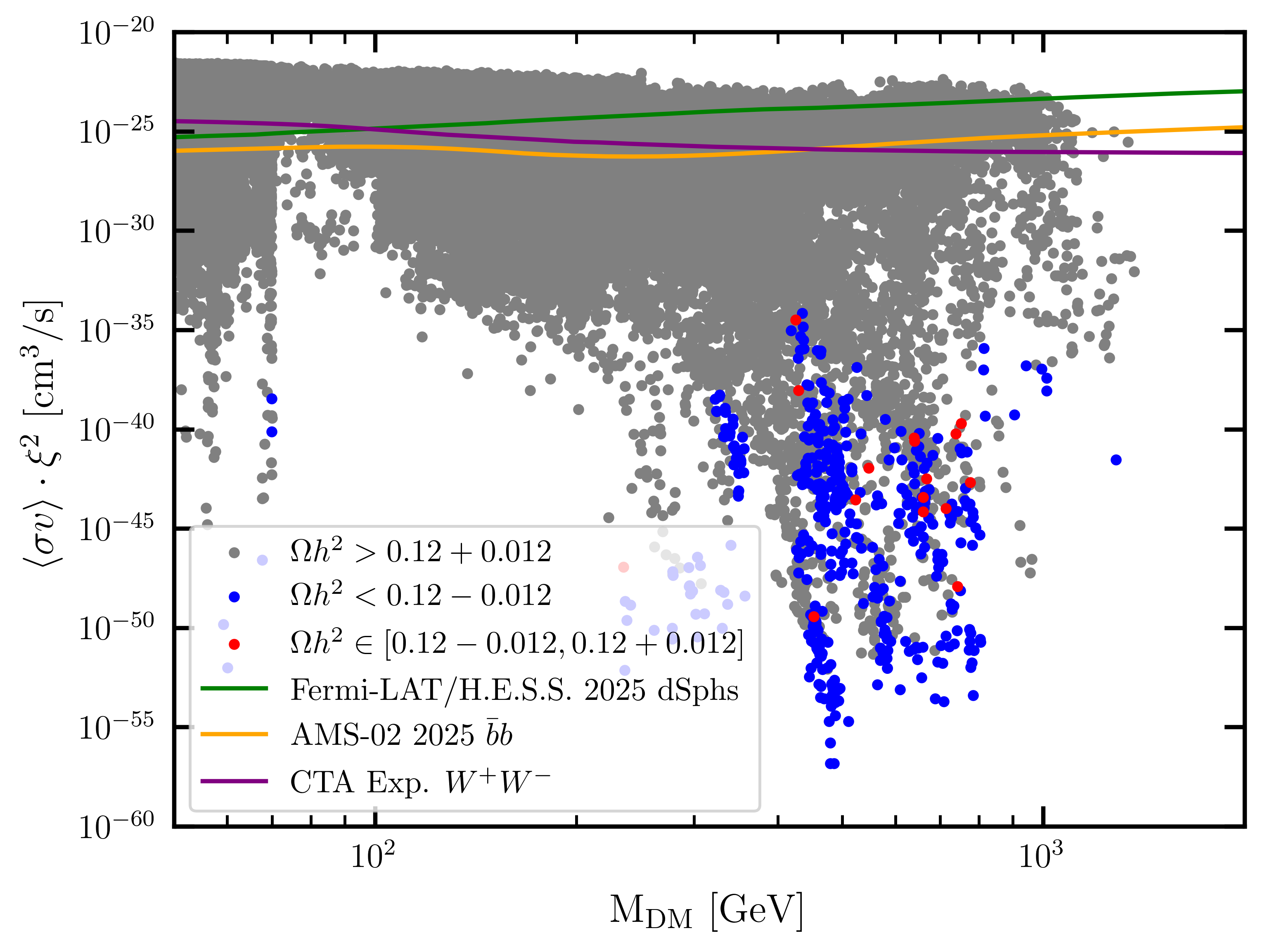}
    \caption{Thermally averaged DM annihilation cross section, $\langle \sigma v \rangle \cdot \xi^2$, as a function of the DM mass. Gray points correspond to overabundant solutions, blue points to underabundant scenarios, and red points denote parameter space consistent with the observed relic density within $3\sigma$. The solid green line shows the combined limits from Fermi-LAT and H.E.S.S. observations of dwarf spheroidal galaxies, while the orange and purple lines represent, respectively, the AMS-02 bound and the projected CTA sensitivity for an Einasto density profile. The latter are included only for reference. All viable solutions lie several orders of magnitude below the current indirect-detection limits, reflecting the strong suppression of the DM annihilation cross section in this model.}
    \label{fig:id_results}
\end{figure}

The results are shown in Fig.~\ref{fig:id_results}, where we present the thermally averaged annihilation cross section as a function of the DM mass. We observe that all viable solutions lie several orders of magnitude below the current limits from Fermi-LAT and H.E.S.S., indicating that indirect detection does not impose significant constraints on the model.

This suppression is mainly driven by the small Yukawa couplings required to satisfy relic density and lepton flavor violation constraints, as well as the rescaling by the DM relic fraction. As a result, the predicted annihilation cross section at present times remains well below the sensitivity of current gamma-ray searches across the entire parameter space.

We also show the limits from the AMS-02 spectrometer only as a qualitative reference~\cite{Reinert:2017aga}. Since AMS-02 constraints are typically derived assuming annihilation into hadronic final states such as $b\bar{b}$ and $W^+W^-$, their direct application to our scenario, where the dominant annihilation channels are leptonic, particularly $\tau^+\tau^-$ and $\mu\tau$, is not straightforward. In particular, mixed-flavor final states such as $\mu\tau$ do not have a direct correspondence with the available experimental limits. Moreover, the interpretation of charged cosmic-ray measurements is subject to significant astrophysical uncertainties associated with cosmic-ray propagation.

For completeness, we also display the projected sensitivity of the Cherenkov Telescope Array (CTA) for observations of the Galactic Center~\cite{CTAConsortium:2017dvg}. Among the gamma-ray searches included in our comparison, this projection provides the strongest expected sensitivity and is obtained for DM annihilation into $W^+W^-$ assuming an Einasto profile for the Galactic DM density. Nonetheless, this curve cannot be interpreted as a direct exclusion limit on our parameter space as it assumes a $100\%$ branching fraction into $W^+W^-$, producing different gamma-ray spectra. Finally, the projected sensitivity depends strongly on the assumed DM distribution in the inner Galaxy, particularly on the choice of the Einasto profile and the corresponding $J$-factor. For these reasons, both the AMS-02 and CTA curves are shown only as qualitative benchmarks, while the robust exclusions applied in our analysis are based primarily on the current gamma-ray constraints from Fermi-LAT and H.E.S.S.

When imposing the compressed spectrum conditions, we observe that the indirect detection behavior remains qualitatively similar to the unrestricted scenario, but with a larger number of points satisfying the relic density constraint. In particular, several regions that were previously overabundant become compatible with the observed DM abundance once the compressed condition is applied.
However, the viable points still predict very suppressed values of $\langle \sigma v \rangle \cdot \xi^2$, lying several orders of magnitude below the current sensitivities from dSphs searches and AMS-02 antiproton measurements. Compared to the case without an upper threshold on $\Delta m$, no significant enhancement of the present-day annihilation cross section is observed. This indicates that, although the compressed spectrum enlarges the viable parameter space through freeze-out effects, it does not substantially improve the indirect detection prospects of the model. Therefore, indirect detection constraints continue to have a negligible impact on the fermionic DM scenario considered here.

\subsection{Collider phenomenology}

\begin{figure*}[t]
    \centering

    \includegraphics[scale=0.7]{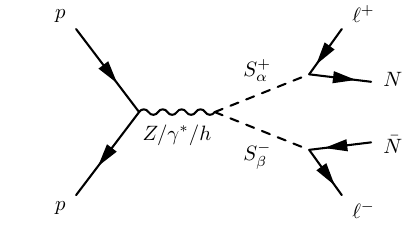}
    \includegraphics[scale=0.5]{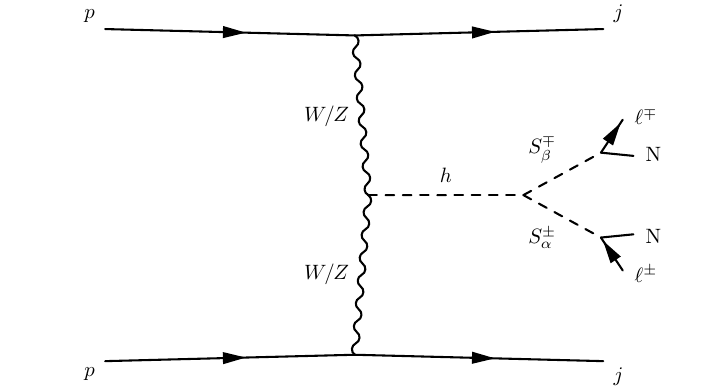}
    \includegraphics[scale=0.7]{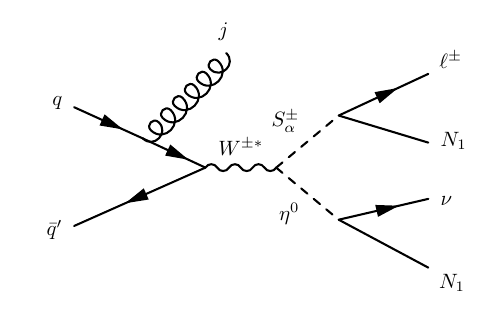}

    \caption{Representative Feynman diagrams illustrating the production of $\textrm{N}_1$ DM in association with soft leptons. Left: sample topology for DY production. Center: sample topology for VBF production. Right: sample topology for DM production accompanied by a mono-jet signal from initial state radiation.}
    \label{fig:feynman_diags}
\end{figure*}

The strong suppression of both direct and indirect detection signals implies that conventional DM searches have limited sensitivity to this scenario. In this context, collider searches provide a complementary and potentially more sensitive probe of the model. The parameter points considered in the following analysis satisfy all the theoretical and experimental constraints discussed in Sec.~\ref{sec:constraints} and have $\Omega h^2$ below the $3\sigma$ upper limit from Planck. Moreover, their predicted direct- and indirect-detection signals remain below the corresponding current experimental limits. We therefore turn to the analysis of their collider phenomenology.

To allow for a direct comparison with collider constraints derived in SUSY compressed spectrum scenarios, we fix the mass splitting to either $\Delta m \leq 30~\textrm{GeV}$ or $\Delta m \leq 50~\textrm{GeV}$. One direct consequence of a compressed mass spectrum in this model is the production of soft SM leptons from the decays $\eta^0\to \textrm{N}_1+\nu_\ell$ and $S_{\alpha}^\mp\to \ell+\textrm{N}_1$. At the LHC, $\textrm{N}_1$ can be produced in association with a soft lepton through either Drell--Yan (DY), mono-jet or vector boson fusion (VBF) processes. The corresponding final states and representative topologies are shown in Fig.~\ref{fig:feynman_diags}.

In VBF processes, the expected experimental signature consists of two forward jets with a large pseudorapidity separation and invariant mass, together with charged leptons or missing transverse energy ($E_T^{\text{miss}}$) in the central region of the detector. This production mechanism has been widely explored in phenomenological analyses as a probe of physics beyond the SM~\cite{Ardila-Tafurth:2025qdf, VBFPhysRevD.87.035029, VBFPhysRevD.90.095022, VBFPhysRevD.91.055025, VBFPhysRevLett.111.061801, natalia2021, qureshi2024probing}, as well as in experimental searches~\cite{Sirunyan2019, CMS:2016ucr, CMS:2015jsu}. In the DY channel, the charged scalars $S^\pm_\alpha$ are produced through an off-shell electroweak gauge boson, giving rise to final states containing charged leptons and $E_T^{\text{miss}}$. The monojet signature, in turn, consists of a single charged lepton, $E_T^{\rm miss}$ generated by the $\mathrm{N}_1$ particles and the neutrino, and a hard initial-state-radiation (ISR) jet recoiling against the remaining decay products.

In the context of scotogenic models, it has been shown in Ref.~\cite{Ardila-Tafurth:2025qdf} that VBF processes can provide an alternative detection channel at high-luminosity colliders such as the High Luminosity LHC (HL-LHC) or the hadronic realization of the Future Circular Collider (FCC-hh)~\cite{Zimmermann:2016puu}. In this analysis, we consider the aforementioned mass splitings $\Delta m \leq 50~\textrm{GeV}$, and $\Delta m\leq 30~\rm GeV$.

Besides producing soft visible final states, the compressed spectrum may also have important consequences for the decay length of the charged scalars. In particular, the reduced phase space available in the decay into the DM candidate can suppress the corresponding decay width and lead to non-prompt signatures. Since the lifetime determines whether the produced charged scalars should be tested using prompt, displaced, or long-lived-particle searches, we first examine their decay lengths before discussing their production cross sections.

\subsubsection{Long-lived charged scalars}

The model naturally accommodates long-lived charged particles (LLPs) as both the couplings and the available phase space entering the charged-scalar decays can be strongly suppressed, a common mechanism for generating LLPs in extensions of the SM~\cite{Alimena:2019zri}. From the Lagrangian given in Eq.~\eqref{eq:lagrangian}  we determine that the dominant decay modes are
\begin{align}
S_i^\pm \rightarrow \textrm{N}_1 \ell_\alpha^\pm,
\end{align}
where $\textrm{N}_1$ is the DM candidate and $\ell_\alpha=e,\mu,\tau$. As these scalars are the result of a mixing, their corresponding effective couplings depend on the Yukawa interactions and on the singlet-doublet composition of each charged state. Consequently, small Yukawa couplings, together with a suppressed doublet or singlet component, can considerably reduce the decay width.

Neglecting the charged-lepton mass, the partial decay width is approximately given by
{
\begin{align}
\Gamma(S_i^\pm \to \textrm{ N}_1\ell_\alpha^\pm)
\simeq
\frac{m_{S_i^\pm}}{16\pi}
\left(
\lvert g^L_{i\alpha 1}\rvert^2
+
\lvert g^R_{i\alpha 1}\rvert^2
\right)
\left(
1-\frac{m_{\textrm{N}_1}^2}{m_{S_i^\pm}^2}
\right)^2 \, ,
\end{align}
}
where $g^{L,R}_{i\alpha 1}$ contain the relevant Yukawa and charged-scalar mixing factors. The total width is obtained by summing over the kinematically accessible lepton flavors.

In the compressed regime, the phase-space factor introduces an additional suppression,
\begin{align}
\left(
1-\frac{m_{\textrm{N}_1}^2}{m_{S_i^\pm}^2}
\right)^2
\propto
\frac{\Delta m_i^2}{m_{S_i^\pm}^2},
\qquad
\Delta m_i
\ll m_{S_i^\pm}.
\end{align}
Thus, even when the two-body decay remains kinematically open, the combined coupling and phase-space suppression may produce a small total width and a macroscopic proper decay length,
\begin{align}
c\tau_{S_i^\pm}=
\frac{\hbar c}{\Gamma_{S_i^\pm}}.
\end{align}

The charged-scalar lifetime is therefore essential for determining the relevant collider signature. Depending on $c\tau_{S_i^\pm}$, the production processes
\begin{align}
pp\rightarrow S_i^+S_i^-,
\qquad
pp\rightarrow S_i^\pm\eta_{R/I},
\end{align}
may give rise to prompt leptons and missing transverse momentum, displaced leptons, kinked tracks, or detector-stable charged-particle signatures~\cite{Alimena:2019zri,ATLAS:2024vnc, ATLAS:2025fdm}. Since these signatures are targeted by different experimental searches, the applicability of a prompt collider analysis cannot be established from the charged-scalar mass alone.

\begin{figure}[H]
\centering
\includegraphics[width=\columnwidth]{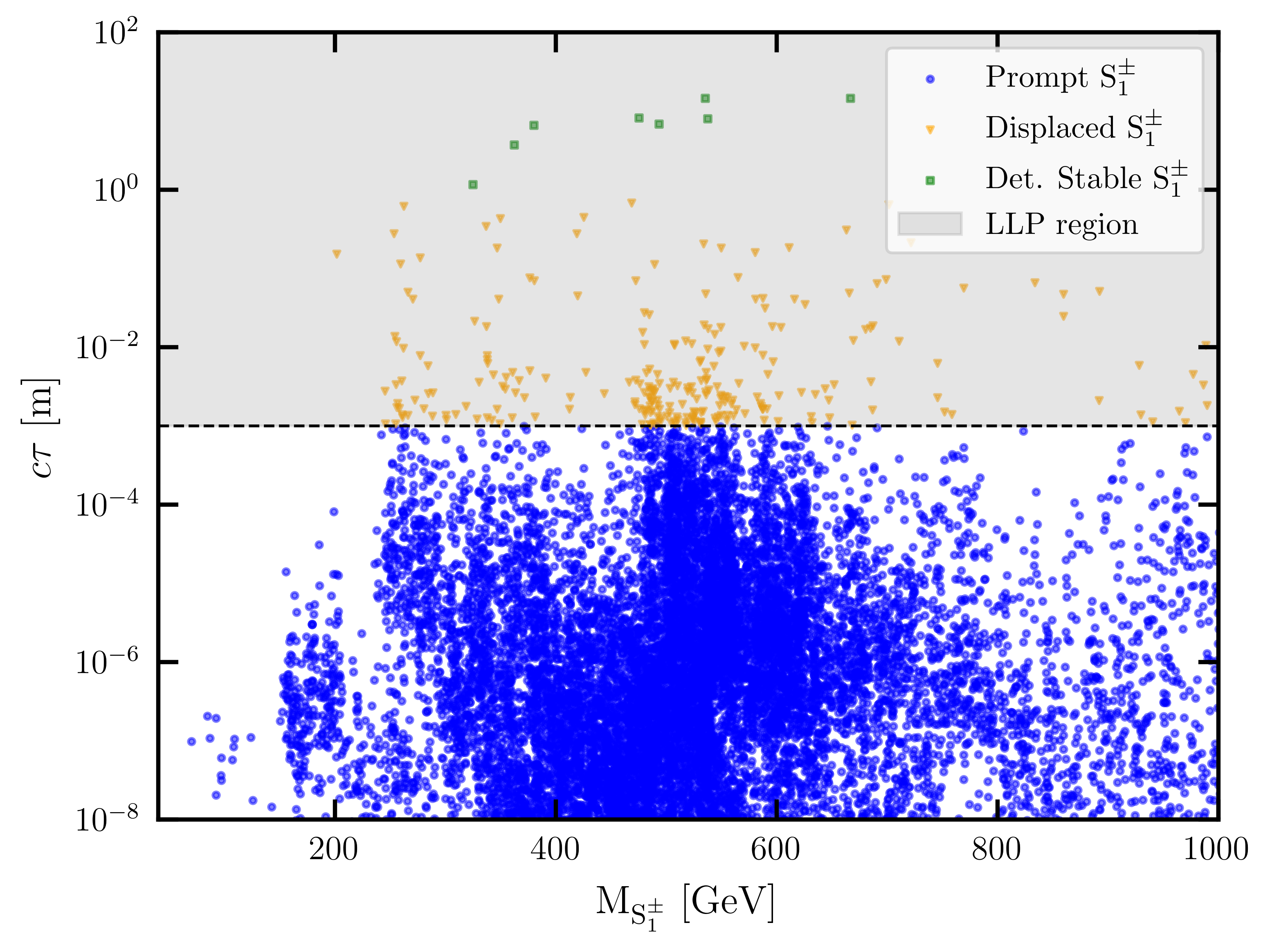}
\caption{Proper decay length of $S_1^\pm$ versus its mass. Prompt ($c\tau<1~\mathrm{mm}$), displaced ($1~\mathrm{mm}\leq c\tau<1~\mathrm{m}$), and detector stable ($c\tau\geq1~\mathrm{m}$) regimes are indicated. The shaded non-prompt region is not excluded and requires a dedicated LLP reinterpretation.}
\label{fig:lifetimes_hhp1}
\end{figure}

Figure~\ref{fig:lifetimes_hhp1} shows the proper decay length of $S_1^\pm$, which is the longest-lived charged scalar throughout the viable parameter space considered here. It therefore provides the most conservative test of whether charged-scalar decays can be described by the prompt collider analyses presented below. Since these analyses rely on the standard reconstruction of particles originating near the primary interaction vertex, we require
\begin{align}
c\tau_{S_1^\pm}<1~\mathrm{mm}.
\label{eq:prompt-decay-condition}
\end{align}
Parameter points satisfying Eq.~\eqref{eq:prompt-decay-condition} are retained for the prompt collider analysis. For larger decay lengths the experimental signature changes qualitatively:
\begin{align}
1~\mathrm{mm}\leq c\tau_{S_1^\pm}<1~\mathrm{m}
\end{align}
may produce displaced decays or kinked tracks, while
\begin{align}
c\tau_{S_1^\pm}\geq1~\mathrm{m}
\end{align}
may result in charged particles traversing a significant fraction of the detector. These boundaries serve only to separate the expected phenomenological regimes: the non-prompt points are not discarded as unviable, but fall beyond the scope of the prompt searches considered here and would require a dedicated reinterpretation in terms of displaced-decay or heavy-stable-charged-particle searches~\cite{ATLAS:2025fdm, ATLAS:2024vnc}. Restricting the analysis to Eq.~\eqref{eq:prompt-decay-condition} ensures that the event-simulation and object-reconstruction assumptions employed below remain consistent with the predicted charged-scalar lifetimes.

Having identified the parameter region in which the charged scalars decay promptly, we now turn to their production at hadron colliders. In this regime, the resulting signatures can be consistently confronted with searches based on standard lepton and missing-transverse-momentum reconstruction. The sensitivity of these searches is then primarily determined by the charged-scalar production rates, which depend on their electroweak interactions and singlet--doublet composition.
\subsubsection{Drell--Yan and VBF searches}
\label{sec:DY_VBF}
For masses around $150~\textrm{GeV}$, the cross sections obtained with \texttt{MadGraph} are of order $\mathcal{O}(10^{-5})~\textrm{pb}$ for the DY channel and $\mathcal{O}(10^{-7})~\textrm{pb}$ for VBF. These small production rates can be understood from the structure of the signal processes, which involve two intermediate charged scalars, $S_i^\pm$. As these physical states arise from the mixing described in Eq.~\eqref{eq:mixing_chargedScalars}, their couplings to the electroweak gauge bosons, the Higgs boson, and the fermionic sector of the model explicitly depend on the corresponding mixing angle. For instance, the $Z S_i^\pm S_j^\mp$ coupling is given by
\begin{equation}
g^Z_{ij}=
\frac{e}{s_Wc_W}
\begin{pmatrix}
\dfrac{1}{2}\cos^2\theta-s_W^2 & -\dfrac{1}{4}\sin2\theta \\
-\dfrac{1}{4}\sin2\theta & \dfrac{1}{2}\sin^2\theta-s_W^2
\end{pmatrix}_{ij},
\end{equation}
where $s_W=\sin\theta_W$ and $c_W=\cos\theta_W$. Thus, the $Z$-boson interaction probes the doublet component of the charged scalar states, while the off-diagonal coupling is directly controlled by $\sin2\theta$. Analogous mixing-dependent structures appear in the $W^\pm$ and Higgs interactions. In contrast, the photon coupling is fixed by electric charge and remains diagonal in the physical basis,
\begin{equation}
g^\gamma_{ij}=e \, \delta_{ij} \, .
\end{equation}

Therefore, the suppression of the complete DY signal does not necessarily
originate solely from the gauge-production vertex. Although the diagonal
photon-mediated production of charged scalars is fixed by their electric
charge, the full signal topology,
$pp\to S_\alpha^+S_\beta^-\to \ell_j^+N_i\,\ell_l^-N_k$, also depends on
the couplings governing the decay of the two charged-scalar legs.
Schematically, the corresponding amplitude can be written as
\begin{equation}
\mathcal{M}_{\alpha \beta}\sim \frac{g^{V}_{S_\alpha^+S_\beta^-} \,\, g_{S_\alpha\ell N} \,\, g_{S_\beta\ell N}}
     {\left(p_\alpha^2-m_{S_\alpha}^2+i m_{S_\alpha}\Gamma_{S_\alpha}\right)\left(p_\beta^2-m_{S_\beta}^2+i m_{S_\beta}\Gamma_{S_\beta}\right)} \, ,
\end{equation}
where $V=\gamma,Z$ and $g_{S_\alpha\ell N}$ denotes the coupling
between the charged scalars and the fermions of the model. In the
convention adopted here, these couplings are
\begin{align}
g_{S_1N_i\ell_j}
    &= i\cos\theta\,y_{ij},
&
g_{S_2N_i\ell_j}
    &= -i\sin\theta\,y_{ij},
\end{align}
with the corresponding Hermitian-conjugate couplings implied for the
charge-conjugate decay. Neglecting the remaining kinematic factors, the
different charged-scalar contributions therefore scale as
\begin{align}
\mathcal{M}_{11}
    &\propto
    g^{V}_{S_1^+S_1^-}\cos^2\theta\,
    y_{ij}y_{kl}^{*} \, ,\\\nonumber
\mathcal{M}_{12,21}
    &\propto
    g^{V}_{S_1^+S_2^-}\sin\theta\cos\theta\,
    y_{ij}y_{kl}^{*} \, ,
\\\nonumber
\mathcal{M}_{22}
    &\propto
    g^{V}_{S_2^+S_2^-}\sin^2\theta\,
    y_{ij}y_{kl}^{*} \, .
\end{align}
Thus, all contributions contain the product of two Yukawa couplings,
one from each charged-scalar decay leg. The $S_1^+S_1^-$ contribution
is therefore Yukawa suppressed, although it does not receive an
additional small-mixing-angle suppression for $\theta\ll1$, since
$\cos\theta\simeq1$. Contributions involving one or two $S_2^\pm$
states acquire, respectively, one or two additional powers of
$\sin\theta$ at the amplitude level. For Yukawa couplings of a common
characteristic size $y$, the corresponding squared amplitudes scale,
at the level of the numerator, as
\begin{equation}
|\mathcal{M}_{11}|^2\propto y^4,\quad
|\mathcal{M}_{12,21}|^2\propto y^4\theta^2,\quad
|\mathcal{M}_{22}|^2\propto y^4\theta^4,
\end{equation}
before accounting for any additional mixing dependence of the production
vertex. 

Since the parameter space probed by the scan is characterized by small
mixing angles, $\theta\simeq0.01$--$0.1~\textrm{rad}$, the charged-scalar
sector approaches a mixing-suppressed regime in which interactions
involving $S_2^\pm$ and off-diagonal charged-scalar transitions are
strongly reduced, while the $S_1^\pm$ productions remain Yukawa suppressed. The same mechanism affects the VBF channel, whose
production amplitude can contain additional mixing-dependent vertices,
such as $hS_\alpha^\pm S_\beta^\mp$, while the two charged-scalar decays
introduce the same $S_\alpha^\pm\ell \textrm{N}_1$ coupling or, in the on-shell
regime, the corresponding branching-ratio factors. Contributions
involving mixed or $S_2^\pm$-dominated states can therefore acquire
several powers of the small mixing angle and become more strongly
suppressed than the corresponding $S_1^+S_1^-$ contribution. The
combination of these mixing suppression, the Yukawa structure, and the
relevant leptonic branching ratios helps explain why the predicted cross
sections are considerably smaller than those obtained in the dynamical
scotogenic model of Ref.~\cite{Ardila-Tafurth:2025qdf}, as well as in
SUSY scenarios in which the production rates are predominantly controlled
by unsuppressed electroweak gauge interactions~\cite{SUSYCMS2019}.
Nevertheless, VBF topologies remain potentially competitive probes of
other BSM scenarios, including supersymmetric models
\cite{SUSYCMS2019,Dutta:2012xe,Dutta:2014jda} and $Z'$ models
\cite{Florez:2016uob,CMS:2024vhy}, because their characteristic
kinematic features---such as two forward jets with a large
pseudorapidity separation and a large invariant mass---can substantially
suppress the relevant SM backgrounds.

It should be stressed, however, that the $y^4$ scaling above refers to the off-shell part of the amplitude. Over most of the phase space the charged scalars are produced on their mass shell, as in resonant pair production $pp\to S_i^+S_i^-$, where the cross section is fixed by unsuppressed gauge interactions. Since $\lambda_4>0$ throughout our scan keeps the decay $S_i^\pm\to\eta_{R/I}\,W^{\pm}$ kinematically closed, the channel $S_i^\pm\to \textrm{N}_1\,\ell^\pm$ is the only one available and the associated branching ratio is exactly one, independently of the size of $y$. Since $S_1^\pm$ is predominantly doublet-like, with $\cos\theta\approx1$, its resonant pair-production rate is not suppressed by the charged-scalar mixing angle. The small effective leptonic couplings and the compressed spectrum instead reduce the decay width, causing a significant fraction of the produced charged scalars to fall outside the prompt-decay category. The lack of prompt DY and VBF events in our scan is therefore driven by the lifetime cut of Eq.~\eqref{eq:prompt-decay-condition}. The smallness of $y$ in our sample should moreover be understood as a consequence of the LFV constraints acting on this Yukawa matrix. However, as in the minimal scotogenic scenario, isolated regions with sizable $y$ compatible with $\mathrm{BR}(\mu\to e\gamma)$ can be found through a dedicated tuning of the flavor structure, although at the cost of significant tension~\cite{Vicente:2014wga}. Since they are strongly disfavored by the global likelihood, they are not populated by our MCMC, which preferentially samples the more natural small-$y$ regions.

As the predicted DY and VBF production rates considered here lie well below the expected sensitivity of compressed-spectrum searches, even at the HL-LHC, probing these channels is particularly challenging. We stress, however, that this conclusion follows from the suppressed production rates characteristic of this model and should not be interpreted as a general statement about DY or VBF searches in other BSM scenarios. Moreover, it does not necessarily extend to other collider topologies within the present model. In particular, monojet searches, in which the invisible system recoils against a hard initial-state-radiation jet, may retain sensitivity to part of the parameter space and will therefore be considered separately.

\subsubsection{Mono-jet searches}

Building on the previous DY and VBF results, we now focus on the monojet topology associated with the production process
\begin{align}
p~p \rightarrow \ell^\pm +\textrm{N}_1 +\textrm{N}_1+ \nu_\ell + j,
\end{align}
as taken from the Feynman Diagrams given in Fig.~\ref{fig:feynman_diags}. In this process, the hard jet is predominantly generated through initial-state radiation and provides the recoil against the weakly interacting final state. The two DM particles, together with the neutrino, generate the missing transverse momentum, while the charged lepton constitutes the only additional visible decay product.

To assess the experimental sensitivity for this topology, we compare our predicted cross sections with the limits reported by the CMS collaboration in Ref.~\cite{CMS:2019zmn}, based on a phenomenological study performed in Ref.~\cite{Florez:2016lwi}. This search targets compressed supersymmetric spectra, through events contaning a soft $\tau$ lepton, a jet with high transverse momentum from initial state radiation, and large missing transverse energy. In particular, we use the limits obtained for direct production of left-handed stau pairs in association with ISR, followed by the $\tilde{\tau}\to \tau+\chi_1^0$ decay. Owing to the similarity between this experimental signature and the compressed topology considered here, these results provide a useful benchmark for estimating the sensitivity to our model. Nevertheless, the comparison is performed only at the cross-section level, since differences in the production mechanisms, particle spins, decay kinematics, and signal acceptance prevent a direct application of the full exclusion limits.

The monojet topology is particularly relevant in the compressed regime, where the charged lepton is expected to have a relatively low transverse momentum and may therefore fail the reconstruction and trigger requirements. For this reason, we do not consider the corresponding pure single-lepton topology without the accompanying hard jet, since its sensitivity would be strongly limited both by the softness of the lepton and by the transverse-momentum thresholds imposed by the experimental triggers.

\begin{figure}[H]
    \centering
    \includegraphics[width=\columnwidth]{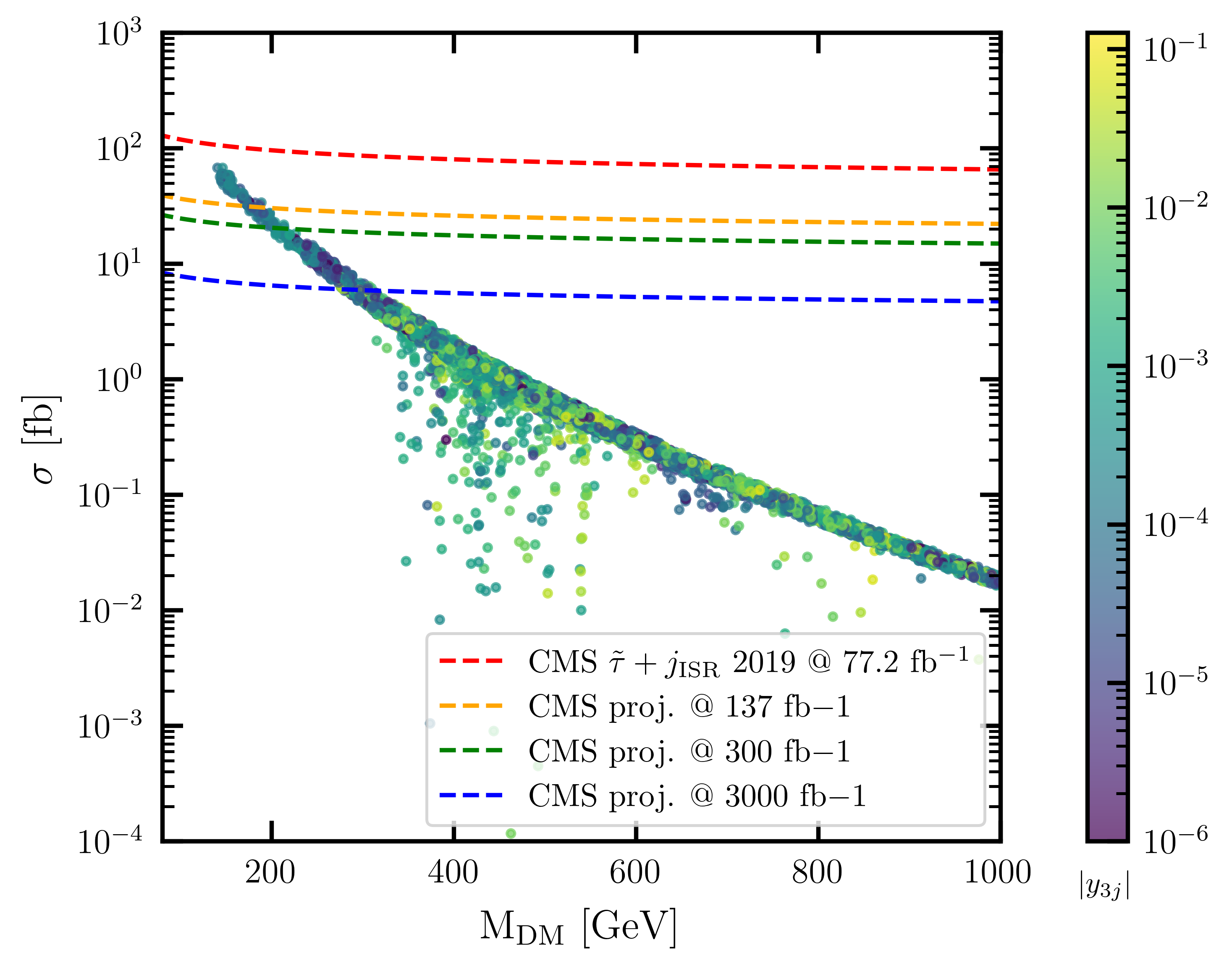}
    \caption{Predicted monojet production cross section as a function of the DM mass, $M_{\rm DM}$, for the viable parameter points of the charged-scalar model. The colour scale indicates the magnitude of the Yukawa coupling $|y_{3j}|$. The dashed curves show the current CMS sensitivity obtained with $77.2~{\rm fb}^{-1}$ and its projections to integrated luminosities of $137$, $300$, and $3000~{\rm fb}^{-1}$. Parameter points lying above a given curve are expected to be within the corresponding experimental sensitivity. While the present CMS result does not constrain the model, part of the low-mass region could be probed with the full Run-2 dataset and at the HL-LHC. These curves are shown as a cross-section-level comparison; a dedicated recast would be required to derive a formal exclusion.}
    \label{fig:monojet_xs}
\end{figure}

To assess the monojet sensitivity to the model, Fig.~\ref{fig:monojet_xs} compares the
predicted production cross section as a function of the DM mass,
$m_{\textrm{N}_1}$, with the CMS cross-section limits reported in
Ref.~\cite{SUSYCMS2019}, obtained with an integrated luminosity of
$77.2~\mathrm{fb}^{-1}$. The color scale indicates the magnitude of the
Yukawa coupling $\left|y_{j3}\right|$. The current CMS sensitivity lies
above essentially the entire viable parameter space and therefore does not
presently constrain the model. However, the projected sensitivities for
larger integrated luminosities indicate that part of the viable parameter
space could become accessible, particularly at the HL-LHC.

The CMS limit used for this comparison is presented as a function of the
chargino mass, $m_{\widetilde{\chi}_1^\pm}$, assuming the fixed mass
splitting
\begin{equation}
    \Delta m_{\rm CMS}
    =
    m_{\widetilde{\chi}_1^\pm}
    -
    m_{\widetilde{\chi}_1^0}
    =
    50~\mathrm{GeV}.
\end{equation}
To express the CMS sensitivity on the same horizontal axis as our
predictions, we rewrite the corresponding mass coordinate as
\begin{equation}
    m_{\widetilde{\chi}_1^0}
    =
    m_{\widetilde{\chi}_1^\pm}
    -
    \Delta m_{\rm CMS},
\end{equation}
and identify $m_{\widetilde{\chi}_1^0}$ with $m_{\textrm{N}_1}$ solely for the
purpose of comparing the two mass axes. This identification should not be
confused with the compressed-spectrum condition imposed in our model,
\begin{equation}
    \Delta m
    =
    \left|m_{\textrm{N}_1}-m_{\eta_I}\right|,
\end{equation}
which involves different states and does not enter the conversion of the
CMS mass axis.

The CMS result covers chargino masses only up to approximately
$500~\mathrm{GeV}$, corresponding to neutralino masses up to approximately
$450~\mathrm{GeV}$ for the assumed value of $\Delta m_{\rm CMS}$. Since
Fig.~\ref{fig:monojet_xs} extends up to $m_{\textrm{N}_1}=1~\mathrm{TeV}$, the sensitivity shown beyond
this range is obtained by extrapolating the mass dependence of the
published upper-limit curve beyond its last reported mass point.

To estimate the potential reach of larger datasets, the corresponding
cross-section sensitivity is further projected according to
\begin{equation}
    \sigma_{\mathrm{lim}}(\mathcal{L})
    =
    \sigma_{\mathrm{lim}}(\mathcal{L}_0)
    \sqrt{\frac{\mathcal{L}_0}{\mathcal{L}}},
\end{equation}
where $\mathcal{L}_0=77.2~\mathrm{fb}^{-1}$. This rescaling assumes that
the signal acceptance and selection efficiency, background composition,
and treatment of systematic uncertainties remain unchanged with
increasing luminosity. It also assumes that the extrapolated mass
dependence of the CMS limit remains valid beyond the published mass
range. The resulting curves should therefore be interpreted as
optimistic, statistically dominated cross-section-level estimates rather
than official CMS projections.

Under these assumptions, the projected sensitivities for integrated luminosities of $137$, $300$, and $3000~\mathrm{fb}^{-1}$ intersect the upper envelope of the predicted cross sections. The full Run-2 dataset could begin to probe the lowest-mass region, while the HL-LHC could test viable scenarios with DM masses extending approximately up to $M_{\rm DM}\sim 300~\mathrm{GeV}$. Parameter points lying above a given projected curve would therefore fall within the corresponding estimated experimental sensitivity. For larger DM masses, the rapid decrease of the production cross section places the viable parameter space below even the optimistic HL-LHC projection.

A noticeable spread toward smaller cross sections is observed at intermediate DM masses. These suppressed points reflect the dependence of the signal rate on the detailed scalar-mixing and interaction structure, beyond its dominant mass dependence. Monojet searches therefore emerge as the most promising collider probe among the channels considered, although their sensitivity remains restricted mainly to the low-mass region.

We emphasize that these projections constitute only a cross-section-level estimate. In particular, possible changes in detector performance, background uncertainties, trigger efficiencies, and event-selection acceptance at larger luminosities are not included. Moreover, since the CMS limits were derived for a different underlying signal model, a dedicated recast including the corresponding kinematic distributions and detector response would be required to establish a robust exclusion or discovery reach.

\section{Summary} \label{sec:summ}

In this work, we have performed a comprehensive phenomenological analysis of the complex scoto-singlet model, considering the lightest Majorana fermion, $\textrm{N}_1$, as the DM candidate. This framework extends the minimal scotogenic model with a charged scalar singlet that mixes with the charged component of the inert scalar doublet. The resulting scalar sector introduces new interactions and coannihilation channels while preserving the one-loop mechanism responsible for neutrino mass generation.

We explored the parameter space using a Markov Chain Monte Carlo analysis and imposed a broad set of theoretical and experimental constraints. These include perturbativity, unitarity, boundedness from below, radiative stability of the Higgs mass, electroweak precision observables, neutrino oscillation data, the cosmological bound on the sum of neutrino masses, lepton-flavor-violating processes, Higgs observables, and collider limits from LEP. The surviving parameter points were subsequently used to study the relic density and the direct-, indirect-, and collider-detection prospects of fermionic dark matter.

Our relic-density analysis shows that most of the unrestricted parameter space leads to an overabundant DM population. Nevertheless, viable solutions are obtained in several phenomenologically distinct regions. Near $M_{\textrm{N}_1}\simeq m_h/2$, resonant annihilation through the SM Higgs boson efficiently reduces the relic abundance. Away from the Higgs funnel, viable solutions are mainly associated with coannihilation processes involving the additional scalar states, particularly the gauge-mediated channels
$S_i^\pm S_j^\mp \rightarrow VV,~
\eta^0\eta^0\rightarrow VV$,
where $V=W,Z$. The additional Yukawa interactions involving the charged scalar singlet also increase the annihilation efficiency and allow viable solutions at lower dark matter masses than in the minimal scotogenic scenario.

The relevance of coannihilation becomes especially clear in compressed-spectrum configurations. By requiring a small mass splitting between $\textrm{N}_1$ and the neutral inert scalar, the thermal population of the heavier $\mathbb{Z}_2$-odd states remains significant during freeze-out. Their efficient annihilation and coannihilation channels reduce the final abundance of $\textrm{N}_1$, converting regions that would otherwise be overabundant into viable or underabundant scenarios. Consequently, the compressed spectrum substantially enlarges the allowed DM parameter space.

The direct-detection prospects of the model are considerably less favorable. Since fermionic DM does not interact with quarks at tree level, the leading spin-independent scattering amplitude is generated radiatively. The resulting relic-density-weighted cross sections lie several orders of magnitude below the current LZ bound, and a significant fraction of the viable parameter space falls below the neutrino floor. Imposing the compressed-spectrum condition does not produce a significant enhancement of the scattering rate. Conventional direct-detection experiments are therefore unlikely to probe most of the fermionic DM scenarios considered here.

A similar conclusion is obtained for indirect detection. The present-day annihilation signal is dominated by leptonic final states, particularly $\tau^+\tau^-$ and the mixed-flavor channel $\mu^\pm\tau^\mp$. After accounting for the relative DM abundance, the predicted annihilation cross sections remain several orders of magnitude below the combined Fermi-LAT and H.E.S.S. limits from dSphs. The compressed spectrum modifies the freeze-out dynamics but does not significantly enhance the annihilation rate at present times. The AMS-02 limits and the projected CTA sensitivity were included only as qualitative references because their underlying final-state and astrophysical assumptions do not directly correspond to those of the model.

The strong suppression of the direct- and indirect-detection signals motivates the study of collider signatures. In compressed scenarios, the decays of the additional scalars produce soft charged leptons and missing transverse momentum. Moreover, the simultaneous suppression of the relevant Yukawa couplings and of the available phase space may lead to macroscopic charged-scalar decay lengths. We therefore separated the parameter space into prompt, displaced, and detector-stable regimes according to the proper decay length of $S_1^\pm$, which is the longest-lived charged scalar in the viable sample. Only points satisfying
$c\tau_{S_1^\pm}<1~\mathrm{mm}$
were included in the prompt collider analysis. Points with larger decay lengths are not excluded, but instead require dedicated displaced-object, kinked-track, or heavy-stable-charged-particle searches.

For prompt charged-scalar decays, we studied Drell--Yan, vector-boson-fusion, and monojet production. The DY and VBF cross sections are strongly suppressed by the small singlet--doublet mixing angles preferred by the viable parameter space. The charged scalars approach a decoupling regime in which their effective interactions with the SM electroweak sector become small, placing these production channels below the expected HL-LHC sensitivity.

Among the collider signatures considered, the monojet topology provides the most promising sensitivity. The current CMS result obtained with $77.2~\mathrm{fb}^{-1}$ does not constrain the viable parameter space. We estimated the sensitivity at larger integrated luminosities by rescaling the cross-section limits
assuming unchanged signal acceptance, selection efficiency, background composition, and systematic uncertainties. Under this optimistic, statistically dominated extrapolation, the full Run-2 dataset may begin to probe the lowest-mass region, whereas the HL-LHC could test part of the viable parameter space up to approximately
$M_{\mathrm{DM}}\sim 300~\mathrm{GeV}$.
For larger masses, the rapid decrease of the production cross section places the signal below even the optimistic HL-LHC projection.

We emphasize that this monojet comparison is performed only at the cross-section level. The CMS limits were obtained for a different underlying signal model, and differences in kinematic distributions, trigger efficiencies, detector acceptance, and event-selection efficiencies have not been included. A dedicated recast is therefore required before deriving a robust exclusion or discovery reach.

Overall, the complex scoto-singlet model provides a viable realization of radiative neutrino masses and fermionic dark matter over a broad mass range. Its direct- and indirect-detection signatures are highly suppressed, while compressed spectra generate a rich collider phenomenology involving prompt monojet events as well as displaced and detector-stable charged particles. These searches appear to be the most sensitive prompt collider channel, whereas dedicated long-lived-particle analyses could provide complementary coverage of regions that are inaccessible to conventional searches. A complete detector-level recast of monojet searches, together with a systematic reinterpretation of displaced and heavy-stable-charged-particle limits, constitutes a natural continuation of this work.

\begin{acknowledgements}
The authors express their gratitude to Óscar Zapata and Cristian Rodríguez for insightful discussions on DM phenomenology and the computational implementation of the scan. G.A.T. and A.F. sincerely acknowledge the constant academic and financial support provided for this project by the Faculty of Science at Universidad de los Andes through projects INV-2024-199-3203 and INV-2026-244-4227, as well as by the Fundación para la Promoción de la Investigación y la Tecnología del Banco de la República de Colombia through Convenio No. 202522 – Project No. 5313. V.M.L. and A.V. acknowledge financial support from the Spanish grants PID2023-147306NB-I00, CNS2024-154524 and CEX2023-001292-S (MICIU/AEI/10.13039/501100011033). V.M.L. also acknowledges  financial support from the Spanish grant CNS2023-144124 (MCIN/AEI/10.13039/501100011033 and “Next Generation EU”/PRTR).
\end{acknowledgements}

\bibliographystyle{spphys}
\bibliography{refs.bib}

\end{document}